\documentclass[iop, numberedappendix]{mnras}

\usepackage[pdftex]{graphicx}

\usepackage{amsmath, amsthm, amssymb, amsbsy}
\usepackage{bm}
\let\bibhang\relax
\usepackage[authoryear,round]{natbib}
\usepackage{subfigure}
\usepackage{upgreek}
\usepackage{accents}
\usepackage{color}
\usepackage{comment}
\usepackage{hyperref}
\usepackage{dblfloatfix}
\usepackage[T1]{fontenc}
\usepackage{aecompl}
\usepackage{enumitem}

\newcommand{\jtm}[1]{{\color{red}{[JTM: #1]}}}

\hypersetup{
  colorlinks   = true, 
  urlcolor     = blue, 
  linkcolor    = blue, 
  citecolor   = blue 
}

\newcommand{\prob}{\mathcal{P}}

\newcommand{\invMpch}{h \text{Mpc}^{-1}}

\newcommand{\Rsm}{R_{\rm{sm}}}
\newcommand{\Mturn}{M_{\rm{turn}}}

\newcommand{\Omnow}{\Omega_{\text{m},0}}
\newcommand{\Obnow}{\Omega_{\text{b},0}}

\newcommand{\Qtot}{Q_{\tot}}
\newcommand{\Qint}{Q_{\text{int}}}

\newcommand{\Tcmb}{T_{\gamma}}

\newcommand{\TS}{T_{\text{S}}}

\newcommand{\TR}{T_{\text{R}}}
\newcommand{\Ts}{T_{\text{S}}}
\newcommand{\Tk}{T_{\text{K}}}

\newcommand{\delC}{\delta_{\varphi}}
\newcommand{\contrast}{\varphi}

\newcommand{\tot}{\text{tot}}

\newcommand{\xHIavg}{\overline{x}_{\text{H } \textsc{i}}}
\newcommand{\xHI}{x_{\text{H } \textsc{i}}}
\newcommand{\xHII}{x_{\text{H } \textsc{ii}}}

\newcommand{\dTb}{\delta T_b}

\title[Phenomenological 21-cm models]{A galaxy-free  phenomenological model for the 21-cm power spectrum during reionization}
\author[Mirocha et al.]{
Jordan Mirocha$^1$,\textsuperscript{\thanks{jordan.mirocha@mcgill.ca}}
Julian B.~Mu\~{n}oz$^2$,
Steven R. Furlanetto$^3$,
Adrian Liu$^1$, and \newauthor \space
Andrei Mesinger$^4$ \\
$^1$Department of Physics \& McGill Space Institute, McGill University, 3600 Rue University, Montr\'eal, QC, Canada H3A 2T8 \\
$^{2}$Center for Astrophysics | Harvard \& Smithsonian, 60 Garden St, Cambridge, MA, 02138, USA \\
$^3$Department of Physics and Astronomy, University of California, Los Angeles, CA 90095, USA\\
$^4$Scuola Normale Superiore, Piazza dei Cavalieri 7, I-56126 Pisa, Italy
}

\begin{document}

\pagerange{\pageref{firstpage}--\pageref{lastpage}} \pubyear{2020}
\maketitle

\begin{abstract}
Upper limits from the current generation of interferometers targeting the 21-cm signal from high redshifts have recently begun to rule out physically realistic, though still extreme, models of the Epoch of Reionization (EoR). While inferring the detailed properties of the first galaxies is one of the most important motivations for measuring the high-$z$ 21-cm signal, they can also provide useful constraints on the properties of the intergalactic medium (IGM). Motivated by this, we build a simple, phenomenological model for the 21-cm  power spectrum that works directly in terms of IGM properties, which bypasses the computationally expensive 3-D semi-numerical modeling generally employed in inference pipelines and avoids explicit assumptions about galaxy properties. The key simplifying assumptions are that (i) the ionization field is binary, and composed of spherical bubbles with an abundance described well by a parametric bubble size distribution, and (ii) that the spin temperature of the ``bulk'' IGM outside bubbles is uniform. Despite the simplicity of the model, the mean ionized fraction and spin temperature of the IGM recovered from mock 21-cm power spectra generated with \textsc{21cmfast} are generally in good agreement with the true input values. This suggests that it is possible to obtain comparable constraints on the IGM using models with very different assumptions, parameters, and priors. Our approach will thus be complementary to semi-numerical models as upper limits continue to improve in the coming years.
\end{abstract}

\section{Introduction} \label{sec:intro}
The 21-cm background \citep{Madau1997} has long been recognized as a powerful probe of the intergalactic medium (IGM) before cosmic reionization is complete \citep[see reviews by, e.g.,][]{FurlanettoOhBriggs2006,Morales2010,Pritchard2012,LiuShaw2020}. Because ionized regions are transparent at redshifted 21-cm wavelengths, maps of the 21-cm background during reionization will consist of ``holes'' in regions where there are many galaxies, and a mostly neutral ``bulk'' IGM beyond with a 21-cm signal that traces the gas density, temperature, and Lyman-$\alpha$ background intensity. While the ultimate goal is to map this patchy structure in detail with, e.g., the Square Kilometer Array (SKA), the current generation of interferometers are seeking a statistical detection of the 21-cm power spectrum \citep[LOFAR, MWA, HERA, GMRT, LWA;][]{vanHaarlem2013,Tingay2013,DeBoer2017,Paciga2013,Eastwood2019}, while a complementary suite of current and planned experiments \citep{Bowman2010,Singh2017,Burns2017,REACH,Philip2019} are targeting the sky-averaged ``global'' 21-cm signal \citep{Shaver1999}, which traces the average properties of the IGM as a function of redshift.

In the last few years, several experiments have reported upper limits on the power spectrum of 21-cm fluctuations during reionization \citep{Parsons2014,Patil2017,Barry2019,Mertens2020,HERA2021} and the earlier cosmic-dawn era \citep{Eastwood2019,Gehlot2019,Gehlot2020,Garsden2021,Yoshiura2021}. Scenarios in which the bulk IGM is still colder than the cosmic microwave background (CMB) during reionization give rise to the strongest fluctuations and so will be the first models to be tested as upper limits continue to improve \citep[e.g.,][]{Parsons2014,Pober2015,Greig2016}. Similarly, stronger-than-expected 21-cm signals can arise if the cosmic radio background has contributions other than the CMB \citep{Feng2018}, e.g.,  synchrotron emission from accreting black holes \citep{EwallWice2018}, star-forming galaxies \citep{Mirocha2019}, or from decaying particles \citep{Pospelov2018,Fraser2018}. Indeed, constraints from MWA, HERA, and LoFAR disfavour models with negligible X-ray heating at $z \sim 8-9$  or very strong radio backgrounds \citep{Greig2021LOFAR,HERA2021Theory,Ghara2020a,Ghara2021,Greig2021MWA,Mondal2020}. Of course, the recent report of an absorption signal in the sky-averaged spectrum at $z \sim 17$ from EDGES \citep{Bowman2018} requires an even colder IGM \citep{Barkana2018,Munoz2018,Fialkov2018,Kovetz2018,Boddy2018} or a brighter background \citep{Feng2018,EwallWice2018,Fialkov2019,Mirocha2019} than models in $\Lambda$CDM cosmologies generally predict. However, the most stringent power spectrum upper limits from \citet{HERA2021} are derived at sufficiently low redshifts relative to EDGES ($z \lesssim 10$ vs. $z \simeq 18$) that they cannot yet directly address the EDGES controversy~\citep{Hills2018,Singh2019,Sims2020,Bradley2019,Tauscher2020,Singh2021}.

The theoretical interpretations of 21-cm measurements have thus far been guided mostly by semi-numerical models of reionization \citep{Mesinger2011,Santos2010,Fialkov2014,Hutter2018} and other approximate techniques \citep{Thomas2009,Ghara2015} designed to avoid more accurate, but expensive, radiative transfer simulations \citep[see, e.g.,][]{Gnedin2014,Rosdahl2018,Ocvirk2020,Kannan2022}. Though several inference frameworks have emerged to jointly fit 21-cm and other constraints on reionization \citep[e.g.,][]{Greig2016,Mirocha2015,Ghara2018,Mondal2020}, current power spectrum limits are still quite weak, and MCMC fits have thus yet to provide a strong constraint on any individual astrophysical parameter, with the exception of the ratio of X-ray luminosity to SFR in high-$z$ galaxies \citep{HERA2021Theory}, and instead reveal the $\sim 2-3$ dimensional corners of parameter space that are most strongly disfavoured by the data. Of course, as upper limits improve and eventually become detections, constraints on astrophysical and cosmological parameters are expected to be exquisite \citep[e.g.,][]{McQuinn2006,Greig2015,LiuParsons2016,EwallWice2016,Liu2016,Kern2017,Munoz2019}. In the meantime, however, it may be prudent to focus also on simpler derived quantities, e.g., the mean neutral fraction and temperature of the IGM (or ratio of background temperature to 21-cm spin temperature, $T_R / T_S$), which are more directly probed by 21-cm experiments, and may thus be easier to constrain.

In this paper, we present a simple, phenomenological model for 21-cm fluctuations during reionization that abstracts away galaxies and instead works directly in terms of the mean properties of the IGM and the size distribution of ionized regions. The goal is first and foremost to build intuition for the results of more sophisticated semi-numerical models in use today. In addition, our phenomenological approach lets us break key assumptions built-in to physical models, and may thus help gauge the extent to which constraints on the IGM derived from 21-cm power spectra are model-dependent. Our formalism is similar to that of \citet{Furlanetto2004} and  extensions \citep{Paranjape2014,Paranjape2016}, though we do not attempt to model the size distribution of ionized regions using excursion set arguments. Instead, we parameterize it directly, which offers more flexibility than physical models. Our approach is similar in spirit to other recent efforts aimed at building intuition for, and providing a cross-check of, more detailed numerical simulations of reionization \citep[e.g.,][]{Raste2018,Kaurov2016,McQuinn2018,Schneider2021}, but conceptually simpler than each. It is also complementary to efforts to constrain the IGM while abstracting away astrophysical parameters as much as possible \citep[e.g.,][]{Cohen2017,Mason2019,Mirocha2013}.

In order to gain ground analytically, we will assume fully ionized and spherical bubbles with infinitely crisp edges, whose abundance is well described by a bubble size distribution (BSD) function. In reality, the ionization field is more complicated; as bubbles merge with neighbours their morphologies become complex, resulting in an interconnected network of ionized regions when the global ionized fraction is just $\sim 10$\% \citep{Furlanetto2016}. Partial ionization due to small-scale clumping -- which we neglect here -- can affect the BSD and 21-cm power spectrum  \citep[e.g.,][]{Sobacchi2014,Bianco2021}, as could a strong X-ray background, if sources with soft spectra dominate. Despite these known shortcomings of the approach, we forge ahead nonetheless, in order to thoroughly assess the accuracy of analytic models and the prospects for using them to derive meaningful constraints on the high-$z$ IGM. To our knowledge, there has yet to be such an attempt to push any analytic model of 21-cm fluctuations through to parameter inference, and compare its results to those of a semi-numerical model \citep[though see, e.g.,][for more general comparisons of analytic and semi-numeric models]{Santos2008,Schneider2021}.

The structure of the paper is as follows. In \S\ref{sec:methods} we introduce our phenomenological approach to 21-cm fluctuations and present its basic predictions. Then, in \S\ref{sec:21cmfast_mocks}, we compare various components of the phenomenological model to two illustrative \textsc{21cmfast} models, in order to gauge its accuracy and motivate different modeling choices. We present a forecast in \S\ref{sec:fits}, conducted by fitting our model to mock signals created both by the phenomenological model itself as well as \textsc{21cmfast}. We conclude in \S\ref{sec:conclusions}.

\section{Phenomenological Modeling Framework} \label{sec:methods}
\begin{figure*}
\begin{center}
\includegraphics[width=0.98\textwidth]{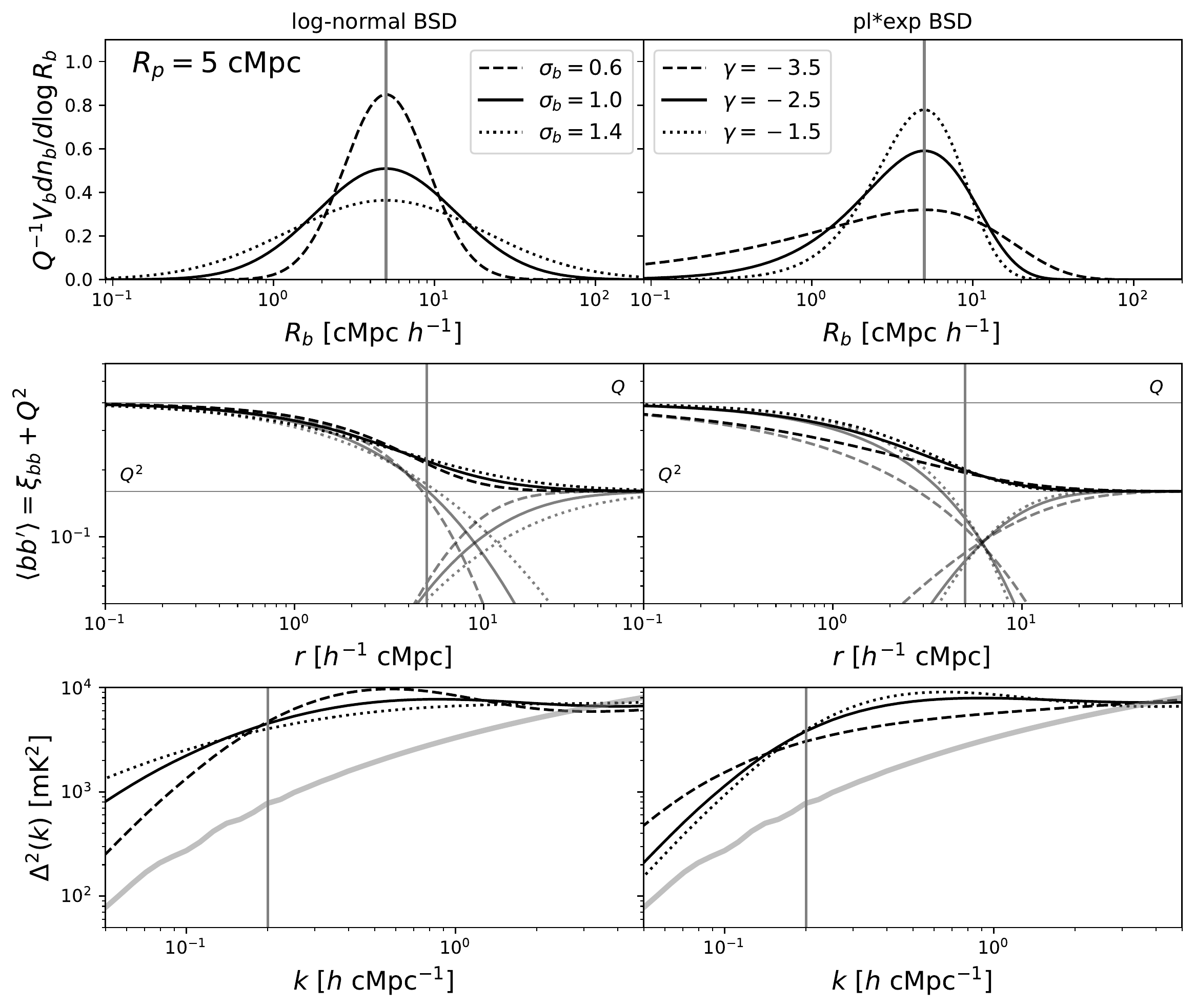}
\caption{{\bf Connection between bubble size distribution (top), bubble correlation function (middle), and 21-cm power spectrum (bottom)}, both for log-normal BSDs (left) and BSDs modeled as a power-law with an exponential cut-off (right). Each model assumes the same mean ionized fraction, $Q=0.2$, and adopts $\Ts = 1.8$ K, as is appropriate for a uniform, adiabatically cooled $z = 8$ IGM. Vertical gray bars in the top and middle rows indicate the location of the peak in the volume-weighted BSD, while in the bottom row we instead show the scale $k_{\mathrm{peak}} = R_{\mathrm{peak}}^{-1}$. Horizontal lines in the middle row indicate the limiting behaviour of the one- and two-bubble terms, which are shown individually in gray. We describe the many terms contributing to 21-cm fluctuations in \S\ref{sec:methods} and Fig. \ref{fig:components}.}
\label{fig:bsd}
\end{center}
\end{figure*}

In this paper, we attempt to remain as agnostic as possible about the source of 21-cm fluctuations at high redshifts. We assume only that the 21-cm field is composed of discrete bubbles embedded in a medium of uniform temperature. We make no effort to model the size distribution of these bubbles via forward modeling, nor do we attempt to evolve the properties of the ``bulk IGM'' beyond bubbles. Instead, we absorb all the astrophysics of reionization and reheating into the size distribution of bubbles, the volume of space they occupy, their mean density, and the mean temperature of the bulk IGM. This section describes the core components of the model, which is implemented in the publicly available \textsc{micro21cm} package\footnote{\url{https://github.com/mirochaj/micro21cm}}. The key predictions of the model are summarized in Figures \ref{fig:bsd},\ref{fig:components}, and \ref{fig:grid2d}, with comparisons to \textsc{21cmfast} and forecasts to follow in \S\ref{sec:21cmfast_mocks} and \ref{sec:fits}.

\subsection{Preliminaries} \label{sec:preliminaries}
The brightness temperature of the 21-cm field at an arbitrary location in space is given by \citep[see, e.g.,][]{Madau1997,FurlanettoOhBriggs2006}
\begin{align}
  \bm{\dTb} & \simeq 27 \ \mathrm{mK} \left(\frac{\Obnow h^2}{0.023} \right) \left(\frac{0.15}{\Omnow h^2} \frac{1 + z}{10} \right)^{1/2} \nonumber \\
    & \times \bm{\xHI} (1 + \bm{\delta})  \left(1 - \frac{\TR}{\bm{\Ts}} \right) \left(\dfrac{H}{\partial_{r} v_{r}}\right), \label{eq:dTb}
\end{align}
where $\xHI = 1 - \xHII$ is the neutral hydrogen fraction, $\delta$ is the baryon density relative to the cosmic mean,
$\partial_{r} v_{r}$ is the line-of-sight gradient of the velocity,
$\TR$ is the temperature of the background (assumed here to be spatially uniform), generally assumed to be the cosmic microwave background $\Tcmb$, and $\Ts$ is the spin temperature, which quantifies the relative abundance of hydrogen atoms in the ground hyperfine triplet and singlet states. Each of these quantities carries an implicit redshift dependence, while bolded quantities are those that vary spatially as well, e.g., $\bm{\dTb} = \bm{\dTb(x)}$, which we will discuss in detail momentarily. In general, $\Ts$ depends the hydrogen and electron densities, gas kinetic temperature, and Lyman-$\alpha$ background intensity, though in this work we abstract away all the physics embedded in $\Ts$ and simply treat it as a homogeneous free parameter. We take the same approach to $T_R$, though in general, e.g., radio emission from galaxies may drive non-trivial fluctuations in $T_R$ as well, and leave interesting signatures in the 21-cm background in some scenarios \citep{Reis2020}.

Given our assumption of a field composed of bubbles, it will be convenient in what follows to rewrite Eq. \ref{eq:dTb} as
\begin{equation}
  \bm{\delta T_b} = T_0(z)  (1 + \bm{\delta}) (1 + \alpha \bm{b}) (1 + \bm{\delC}) \label{eq:dTb_compact}
\end{equation}
where $T_0$ is a redshift- and cosmology-dependent normalization,
\begin{equation}
  T_0 \equiv 27  \left(\frac{\Obnow h^2}{0.023} \right) \left(\frac{0.15}{\Omnow h^2} \frac{1 + z}{10} \right)^{1/2} \left(1 - \frac{\TR}{\Ts} \right) \ \mathrm{mK} . \label{eq:T0}
\end{equation}
and $\delC$ is a fractional perturbation in the temperature `contrast,'
\begin{equation}
    \bm{\contrast} = \frac{\bm{\Ts} - \TR}{\bm{\Ts}} ,
\end{equation}
which is related to fractional perturbations in the spin temperature via
\begin{equation}
    \bm{\delC} = \left(\frac{\TR}{\Ts-\TR} \right) \bm{\delta_{T_S}} = \contrast^{-1} \left(\frac{\TR}{\Ts} \right) \bm{\delta_{T_S}} \label{eq:contrast_pert}
\end{equation}
The variable $\bm{b}$ in Eq. \ref{eq:dTb_compact} represents a binary field of bubbles, and thus takes on values of 0 or 1 only, with 1 indicative of a fully ionized or fully heated bubble. In this work, we will focus entirely on ionized bubbles, noting here the possibility of heated bubbles for completeness -- one can easily switch from one to the other with a suitable choice of $\alpha$, i.e.,
\begin{equation}
\alpha =   \left\{
\begin{array}{ll}
      0 & \text{no bubbles} \\
      -1 & \text{ionized bubbles} \\
      \TR/(\Ts-\TR) & \text{heated bubbles} \\
\end{array} \label{eq:alpha}
\right.
\end{equation}
This expression assumes that heated `bubbles' are fully saturated, $\Ts \gg \TR$, in which case plugging $\alpha=\TR/(\TS - \TR)$ into Eq. \ref{eq:dTb_compact} reduces to the brightness temperature of a saturated patch of the IGM. One could alternatively leave the temperature of heated bubbles as a free parameter.

From this point onward, we will discontinue the use of bold-faced variables, meaning any occurrence of $\Ts$ or $\contrast$ refers to the mean, while $\delta$'s and $b$'s carry all spatial information.

If we assume for the moment that the spin temperature field is spatially homogeneous, leaving $\delC = 0$, we can write the correlation function of the 21-cm field, $\xi_{21} \equiv \langle \delta T_b \delta T_b^{\prime} \rangle - \langle \delta T_b \rangle^2$, relatively compactly as
\begin{align}
  T_0^{-2} \xi_{21} & = \langle \delta \delta^{\prime} \rangle + \alpha^2 \langle b b^{\prime} \rangle - \alpha^2 \langle b \rangle^2 \nonumber \\
  & + 2 \alpha \langle b \delta^{\prime} \rangle  + 2 \alpha^2 \langle b b^{\prime} \delta \rangle + 2 \alpha\langle b \delta \delta^{\prime} \rangle  + \alpha^2 \langle b b^{\prime} \delta \delta^{\prime} \rangle \nonumber \\
  & - 2 \alpha^2 \langle b \rangle \langle b \delta \rangle - \alpha^2 \langle b \delta \rangle^2  \label{eq:xi21_full}
\end{align}
Here, angular brackets indicate ensemble averages and primed quantities indicate points a distance $r$ from unprimed points. So far, our results are exact, as only terms proportional to $\langle \delta \rangle$ have been dropped (the overdensity field has mean zero by definition). The ensemble average of the bubble field is equivalent to the volume filling fraction in this framework, so in what follows we will use $Q \equiv \langle b \rangle$. 

In the next few sub-sections, we will discuss methods for modeling the various terms in Eq. \ref{eq:xi21_full}, as well as 'correction terms' that arise when $\delC > 0$ and peculiar velocities (i.e.~redshift-space distortions) are included. We will adopt the following convention for correction factors that relate an arbitrary fluctuation in $\delta_X$ to the density field, i.e.,
\begin{equation}
    \delta_X \equiv \beta_X \delta . \label{eq:beta_definition}
\end{equation}
In all that follows, we will plot only the dimensionless power spectrum,
\begin{equation}
  \Delta_{21}^2(k) = k^3 P_{21}(k) / 2\pi^2 \label{eq:Delta_sq}
\end{equation}
where $P_{21}$ is related to the 21-cm correlation function by the transform\footnote{We discuss numerical solutions to this integral in \S\ref{sec:dd}.}
\begin{equation}
    P_{21}(k) = \frac{1}{(2\pi)^3} \int 4\pi R^2  \frac{\sin(kR)}{kR} \xi_{21}(R) dR \label{eq:P21}
\end{equation}
Note that it is common in the literature to plot $\overline{\delta T_b}^2 \Delta_{21}^2(k)$, i.e., the global 21-cm signal squared times $\Delta_{21}^2$. The difference is one of definitions and notation. If one works in terms of the fractional perturbation in the 21-cm brightness temperature, such that $\bf{\delta T_b}(\bf{x}) = \overline{\delta T_b} (1 + \delta_{21})$, and defines the 21-cm correlation function as $\xi_{21} \equiv \langle \delta_{21} \delta_{21}^{\prime}\rangle - \langle \delta_{21}\rangle^2$, then the appropriate quantity to plot is indeed $\overline{\delta T_b}^2 \Delta_{21}^2(k)$. However, we work directly in terms of the 21-cm signal's constituent terms, i.e., $\delta$, $b$, $\delta_{\phi}$, etc., which leaves a normalization factor of $T_0^2$ only, where $T_0 = \overline{\delta T_b} (1-Q)^{-1}$ (see also Eq. \ref{eq:T0}). We absorb this factor of $T_0^2$ into the 21-cm correlation function, as in Eq. \ref{eq:xi21_full}, which implicitly lends Equations \ref{eq:Delta_sq} and \ref{eq:P21} units of $\rm{mK}^2$. Hence the absence of $\overline{\delta T_b}^2$ as a multiplicative pre-factor applied to $\Delta{21}^2(k)$ in the $y$ labels of our figures.

\subsection{Adiabatic expansion and compression} \label{sec:adiabatic_corr}
The first correction we consider addresses the correlation between the density and temperature due to adiabatic expansion or compression alone (i.e., no X-ray heating from astrophysical sources). For clarity, we allow $\delC > 0$ but hold $b = 0$, in which case the correlation function of the 21-cm background can be written as
\begin{align}
T_0^{-2} \xi_{21,b=0} & = \langle \delta \delta^{\prime} \rangle +\langle \delC \delC^{\prime} \rangle   \nonumber \\
  & + 2 \langle \delC \delta^{\prime} \rangle  + 2 \langle \delC \delC^{\prime} \delta \rangle + 2 \langle \delC \delta \delta^{\prime} \rangle  + \langle \delC \delC^{\prime} \delta \delta^{\prime} \rangle \nonumber \\
  & - 2 \langle \delC \rangle \langle \delC \delta \rangle - \langle \delC \delta \rangle^2 .
\end{align}
We further assume that density and temperature fluctuations are small, which allows us to eliminate all 3- and 4-pt terms from the above expression, and note that the ensemble average of $\delC$ is zero by construction and can thus be removed as well.

Perturbations in the density, temperature, and ionization can be evolved numerically \citep[e.g.,][]{Naoz2005,Barkana2005,Pritchard2007}. However, \cite{Munoz2015} found that fluctuations in the kinetic temperature  can be more straightforwardly related to $\delta$ from recombination and throughout the cosmic dark ages up to whenever inhomogeneous X-ray heating occurs, via
\begin{equation}
    \delta_T \equiv \dfrac{\delta \Tk}{\Tk} = \beta_T(z) \delta . \label{eq:beta_T}
\end{equation}
The coefficient $\beta_T$, usually termed the adiabatic index, would be exactly 2/3 for pure adiabatic cooling, though the Compton scattering with the CMB produces a deviation from this factor.
For the redshift range of interest, the result in \cite{Munoz2015} can be  approximated as $\beta_T(z) = c_0 - c_1 (z-10)$, with $c_0=0.58$ and $c_1=0.005$ to within 3\% precision (for $z=6-50$).

Using this linear relationship between $\delta$ and $\delta_T$, and assuming saturated Wouthuysen-Field coupling (so that $Ts \approx \Tk$.), we have
\begin{align}
T_0^{-2} \xi_{21,b=0} & = \langle \delta \delta^{\prime} \rangle + \beta_{\varphi}^2 \langle \delta \delta^{\prime} \rangle + 2 \beta_{\varphi} \langle \delta \delta^{\prime} \rangle \nonumber \\
  & = \langle \delta \delta^{\prime} \rangle \left[1 +  \beta_{\varphi} \right]^2
\end{align}
where we've made use of Eq.~\ref{eq:contrast_pert} and defined $\beta_{\varphi} \equiv \beta_T \contrast^{-1} \left(\frac{\TR}{\Ts} \right)$.

The overall effect of adiabatic expansion and compression is to reduce the amplitude of the 21-cm emission in a given patch of the IGM relative to the uniform density case. To see this, we now write Eq. \ref{eq:dTb} with a correction term $\beta_\delta (z) \equiv 1 + \beta_{\varphi}$ applied to the density, i.e.,
\begin{equation}
    \dTb \simeq \overline{T_0} (1 + \beta_\delta \delta),
    \label{eq:Tbdensity}
\end{equation}
which yields
\begin{equation}
    \dTb \simeq  \dfrac{\overline{T_0}}{1-T_\gamma/\Ts} \left[1 - \dfrac{T_\gamma}{T_S} (1-\beta_T)  \right] \delta,
    \label{eq:Tbdensity_bias} .
\end{equation}
As expected, $\beta_T > 0$ reduces the amplitude of the signal, holding all other quantities fixed.

For scenarios where the gas has a temperature above the adiabatic prediction, we ought to account for how heating changes this picture.
We can always write that at every point and $z$,
\begin{equation}
    T_g = T_g^{\rm ad} + \Delta T_g^X,
\end{equation}
where $T_g^{\rm ad}$ is the adiabatic prediction, and $\Delta T_g^X$ is the heating term, which we assume to be approximately homogeneous, as predicted for instance for hard X-rays.
Therefore, $\delta T_g = \delta T_g^{\rm ad} = \beta_T \delta$, and thus
\begin{equation}
    \delta_T = \delta_T^{\rm ad} \dfrac{T_g^{\rm ad}}{T_g} = \beta_T \dfrac{T_g^{\rm ad}}{T_g} \delta,
\end{equation}
or equivalently, we can use our result in Eq. \ref{eq:Tbdensity} with the correction
\begin{equation}
    \beta_T \to  \beta_T \, {\rm min}\left(1,\dfrac{T_g^{\rm ad}}{T_g}\right).
\end{equation}
We note that for $T_g < T_g^{\rm ad}$, as predicted by models of DM-induced cooling \citep[see, e.g.,][]{Munoz2015b,Munoz2018b,Barkana2018,Berlin2018}, our formula predicts an increase in the $\beta_T$ term.
To remain conservative, we cap $\beta_T$ at its adiabatic value of 2/3.


\subsection{Statistics of bubbles} \label{sec:bubbles}
We now turn our attention to 21-cm fluctuations sourced by fluctuations in the ionization field, starting with the auto-correlation term $\langle b b^{\prime} \rangle$ (see Eq. \ref{eq:xi21_full}). Much of this follows from \citet{Furlanetto2004} (hereafter FZH04), but we review it here nonetheless for completeness. Given that $b$ is binary, the ensemble average greatly simplifies,
\begin{equation}
  \langle b b^{\prime} \rangle \equiv \int db \int db^{\prime} b b^{\prime} f(b, b^{\prime})  = \prob_{bb}
\end{equation}
since the integrand is only non-zero when both $b$ and $b^{\prime}$ are unity. In this case, the double integral over the joint distribution $f(b^{\prime}, b^{\prime})$ is simply the probability that two points are both in bubbles, hence our use of the notation $\prob_{bb}$. This probability will be determined entirely by the BSD, which we parameterize flexibly rather than model from physical arguments \citep[see][for excursion set models]{Furlanetto2004,Paranjape2014}. Several example BSDs are shown in Figure \ref{fig:bsd} and discussed in more detail in \S\ref{sec:bsd}.

\defcitealias{Furlanetto2004}{FZH04}

For a field composed of discrete bubbles, $\prob_{bb}$ can be worked out for an assumed size distribution of bubbles, $n_b(R_b)$, following \citetalias{Furlanetto2004}. Drawing inspiration from the halo model \citep{Cooray2002}, only two configurations are possible: either two points are in the same bubble or they are in different bubbles, i.e., $\prob_{bb} = \prob_1 + \prob_2$.

The probability $\prob_1$ that two points are in the same bubble depends on the fraction of the volume in which a single bubble can enclose two points separated by a distance $r$. This amounts to an integral over the bubble size distribution weighted by the ``overlap volume,'' $V_o$,
\begin{equation}
  \prob_1^{\prime}(r) = \int dR_b n_b(R_b) V_o(r, R_b) \label{eq:P_bb1_p}
\end{equation}
The overlap volume is the inter-sectional volume of two spheres of radius $R_b$ separated by $r$,
\begin{equation}
V_o =   \left\{
\begin{array}{ll}
      \frac{4\pi}{3} R_b^3 - \pi r \left[R_b^2 - r^2/12 \right] & r < 2R_b \\
      0 & \text{otherwise} \\
\end{array}
\right.
\end{equation}
A single bubble can only engulf two points separated by $r$ if it is centered within this region.
Note that the probabilities in Eq. \ref{eq:P_bb1_p} (as well as Eq. \ref{eq:P_bb2_p} below) are primed to indicate that they are not the final probabilities used in the model, as corrections are in order (see \S\ref{sec:overlap}).

The other possibility is that two points reside in different bubbles. In this case, we need the probability that a single source can ionize one point but not the other, which is proportional to $V(R_b) - V_o$, with $V(R_b) = 4\pi R_b^3/3$:
\begin{equation}
  \prob_2^{\prime}(r) = (1 - \prob_1^{\prime})
  \times \left( \int dR_b n_b(R_b) \left[V(R_b) - V_o(r, R_b) \right] \right)^2 \label{eq:P_bb2_p}
\end{equation}
The leading factor of $(1-\prob_1^{\prime})$ ensures that the two points under consideration do not reside in a single bubble.

The one- and two-bubble terms are shown individually in the middle row of Fig. \ref{fig:bsd}; the former dominates on small scales, and asymptotes to a value of $Q$, while the latter dominates on large scales, and asymptotes to $Q^2$. On intermediate scales comparable to the typical bubble size, both terms are comparable. The smoothness of the transition from the one-bubble regime to the two-bubble regime is governed by the details of the BSD, which we discuss more in \S\ref{sec:bsd}.

We currently neglect the clustering of bubbles -- a more sophisticated approach may be warranted at early times and/or intermediate scales where clustering of bubbles is important, in which case a factor of $1 + \epsilon$ within one of the integrands of Eq. \ref{eq:P_bb2_p} would be necessary, indicating an excess probability that a second source ionizes point 2 given that the first point is ionized (see, e.g., \S3.3 in \citetalias{Furlanetto2004}).

\subsubsection{Bubble Overlap} \label{sec:overlap}
Perhaps the most obvious shortcoming of the treatment so far is that it neglects the potential for overlap between bubbles. In reality, overlapping bubbles cease to be distinct entities, and will instead form a single larger bubble -- indeed, \citep{Furlanetto2016} showed that throughout most of reionization the vast majority of the ionized volume is contained in a single ``percolating cluster" with a very complex shape (although simulations do suggest that the percolating cluster is composed of subunits with a finite size; e.g.,  \citealt{Lin16, Busch20}). The percolation process qualitatively changes the meaning of the BSD, so that we cannot treat these stages self-consistently. However, motivated by the existence of a characteristic scale in simulations of reionization, we can modify our probabilities slightly to account for overlap in a statistical sense.

To assess the importance of overlap, it is useful to first consider the total volume contained in bubbles,
\begin{equation}
    \Qtot = \int dR_b n_b(R_b) V(R_b)
\end{equation}
This quantity is \textit{not} equivalent to the volume filling fraction of ionized gas, $\Qtot \neq Q$, since there is nothing stopping two (or more) bubbles from co-occupying the same space in our model. It is more accurate to consider $Q$ as the probability that a single point is ionized, which is a sum over all possible configurations, e.g., that a point is engulfed by a single bubble of radius $R_1$, or that a point is instead engulfed by a bubble of radius $R_2$ (but not $R_1$), etc., i.e.,
\begin{equation}
    \prob_b = \prob_1 + (1 - \prob_2) + (1 - \prob_1) (1 - \prob_2) \prob_3 + ...
\end{equation}
Note that each $\prob_i$ term in the above sum requires an integral, and neglects the possibility that a single point resides within multiple bubbles.

One can dramatically simplify this computation by realizing that we only care if a point is ionized, regardless of how many bubbles contain it. The ionized fraction is simply the complement of the probability $P_0$ that \textit{no} bubbles engulf a point, which we can write as a Poisson distribution (see also \S3.1-3.2 in \citetalias{Furlanetto2004}),
\begin{equation}
    Q = 1 - \exp \bigg[-\int dR_b n_b(R_b) V(R_b) \bigg] .  \label{eq:Q}
\end{equation}
The same logic applies to the calculation of one- and two-bubble terms. As a result, we take our final, unprimed probabilities to be
\begin{equation}
  \prob_1(r) = 1 - \exp\bigg[-\int dR_b n_b(R_b) V_o(r, R_b) \bigg] \label{eq:P_bb1}
\end{equation}
and
\begin{align}
  \prob_2(r) & = (1 - \prob_1)
  \nonumber \\
  & \times \left(1 - \exp\bigg[- \int dR_b n_b(R_b) \left(V(R_b) - V_o(r, R_b) \right) \bigg] \right)^2 \label{eq:P_bb2}
\end{align}


To gauge the importance of overlap, we can easily compute the difference of $\Qtot$ and $Q$, i.e., the fraction of the volume composed of \textit{more than one} bubble. We find that this ``global inter-sectional volume,''
\begin{equation}
    \Qint \equiv \Qtot - Q
\end{equation}
is $\ll 1$ when $Q$ is small, indicating that overlap is unimportant at early times, as expected. However, $\Qint$ rises as $Q$ grows, reaching a value of $\Qint \simeq 0.2$ when $Q=0.5$, and $\Qint \simeq Q$ when $Q \simeq 0.8$, i.e., overlap is likely an order unity effect for the last $\simeq 20$\% of reionization. Note that for this calculation we normalize $n_b$ to satisfy a user-supplied value of $Q$ using Eq. \ref{eq:Q}. Note also that $\Qtot$ and $\Qint$ can exceed unity, because these volumes include every occurrence of overlap at a given point, e.g., a single region will contribute three times its volume if three separate bubbles contain it.

\begin{figure*}
\centering
\includegraphics[width=0.98\textwidth]{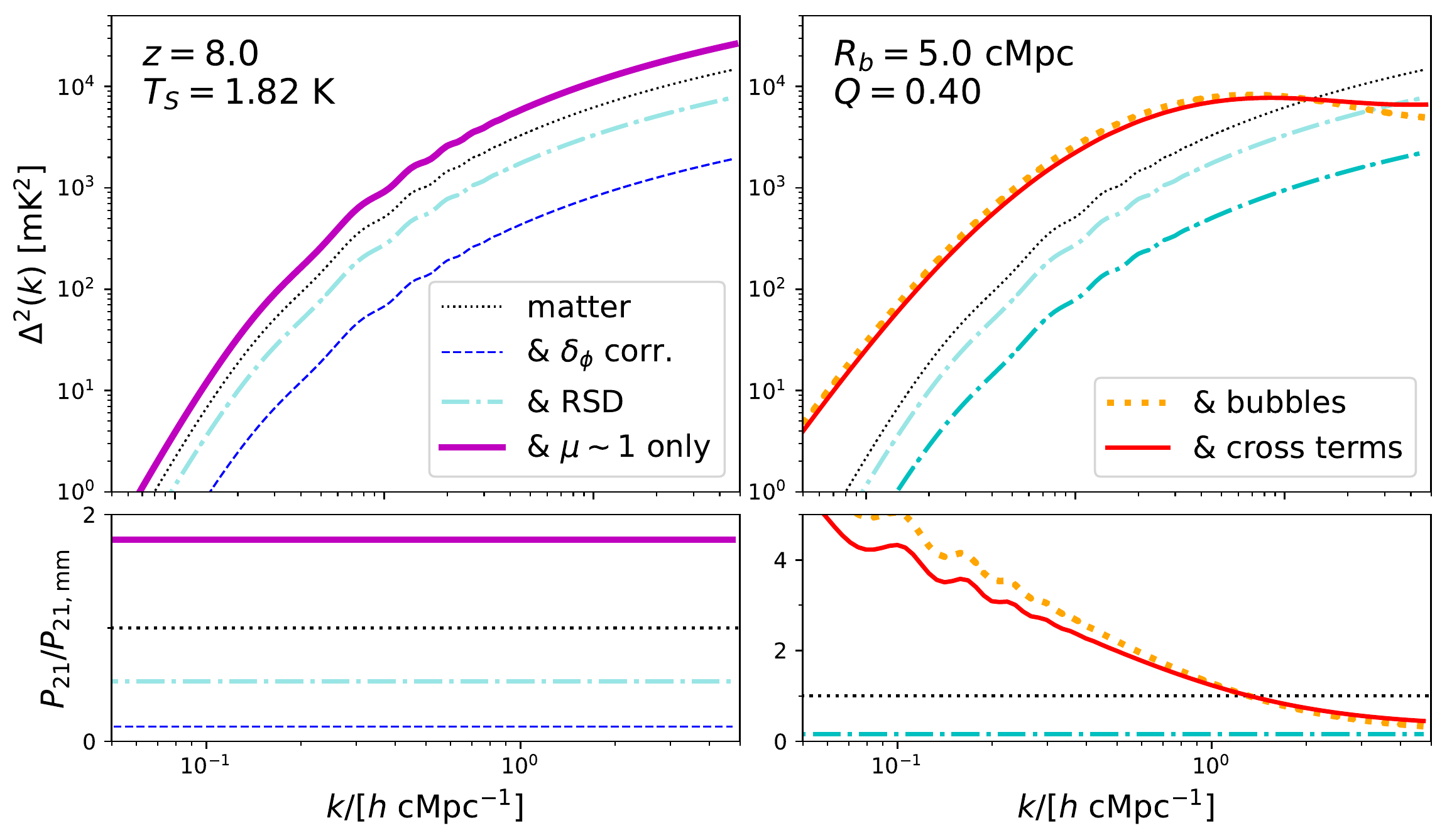}
\caption{{\bf Main contributions to the 21-cm power spectrum.} Starting from a universe with only linear density fluctuations (dotted black; left column), we add corrections for adiabatic compression/expansion (dotted blue), and redshift-space distortions, with spherical averaging (cyan) as well as pure line of sight modes (magenta). Then, in the right-hand column, we add the auto-correlation contributions from bubbles ($\langle bb^{\prime} \rangle$; dotted orange) and the cross-correlation terms involving bubbles and the density field (solid red). The bottom row in each column shows the relative amplitude of the 21-cm power spectrum relative to a model with matter fluctuations only, which we indicate as $P_{21,\rm{mm}}$.
}
\label{fig:components}
\end{figure*}

\subsubsection{Bubble Size Distributions} \label{sec:bsd}
We have yet to explicitly define the BSD, instead leaving it as a generic function $n_b(R_b)$. Rather than model the BSD from a galaxy formation model and the excursion set approach, we parameterize it flexibly\footnote{The parametric forms we choose are known to mimic physically motivated models reasonably well \citep[see, e.g.][]{Kakiichi2017}}. Our fiducial model adopts a log-normal form,
\begin{equation}
    \frac{dn_b}{dR_b} \propto (R_c \sigma)^{-1} \exp\bigg[ -\frac{(R_b - R_c)^2}{2\sigma_b^2}  \bigg], \label{eq:bsd_lognormal}
\end{equation}
peaked near a radius $R_c$, and with a width $\sigma_b$.
We also consider a power-law times an exponential,
\begin{equation}
    \frac{dn_b}{dR_b} \propto \left(\frac{R_b}{R_c}\right)^{\gamma} \exp\bigg[-R_b/R_c \bigg], \label{eq:bsd_plexp}
\end{equation}
with the index $\gamma$ as a free parameter rather than $\sigma$.
In each case, the BSD is normalized to preserve the mean ionized fraction defined in Eq. \ref{eq:Q}.

Note that the more relevant quantity in our analysis is actually the volume-weighted BSD, $V dn_b/d\log R_b$, where $V=4\pi R_b^3 / 3$, since the probability that two points are ionized is related to volume more directly than bubble size. The peak in the volume-weighted BSD is \textit{not} equal to the peak in $dn_b/dR_b$. As a result, in all that follows, we use the variable $R_p$ or $R_{\rm peak}$ to indicate the location of the peak in $V dn_b/d\log R_b$. For both BSDs we explore, it is easy to translate $R_c$ to $R_p$ via
\begin{equation}
    R_p = R_c \exp \{3 \sigma^2\}
\end{equation}
for the log-normal, and
\begin{equation}
    R_p = R_c (\gamma+4)
\end{equation}
for the power-law-times-exponential model.

\subsubsection{Limiting Behaviour Checks}
Before moving on to cross-terms, let us examine the limiting behaviour of the correlation function of the bubble field, $\xi_{bb} \equiv \langle b b^{\prime} \rangle - Q^2$. We have only two obvious requirements: (i) fluctuations must vanish on large scales $r \gg R_b$, and (ii) fluctuations must vanish at $Q=0$ and $Q=1$. First, for an arbitrary value of $Q$, on large scales ($r \gg R_b$) we find $V_o \rightarrow 0$, so $\xi_{bb} \rightarrow \prob_2 - Q^2 \simeq 0$, and fluctuations vanish as they must, since $\prob_2$ tends toward $Q^2$ on large scales. Second, the $Q=0$ limit is satisfied by construction in our framework, since the BSD is normalized by $Q$, thus forcing $\prob_1=\prob_2=0$ when $Q=0$. This leaves only the question of whether or not $\xi_{bb}$ vanishes at $Q=1$.

On large scales, fluctuations vanish regardless of $Q$, as shown above. On small scales, $r \ll R_b$, the overlap volume tends to the bubble volume, $V_o(r, R_b) \rightarrow V(R_b)$. As a result, $\prob_1 \rightarrow Q$, and $\prob_2 \rightarrow 0$. This leaves a bubble correlation function $\xi_{bb} \simeq Q - Q^2 = Q (1 - Q)$, which is indeed zero when $Q = 1$.

\subsection{Cross Correlations Between Ionization and Density} \label{sec:crossterms_2pt}
We have thus far neglected correlations between the density field and bubbles (see second and third rows of Eq. \ref{eq:xi21_full}). In keeping with the phenomenological spirit of this paper, we take a simple approach to these terms that abstracts away assumptions about the sources (to the extent that this is possible). The busy reader may skip ahead to Fig.~\ref{fig:components}, which shows the effects of cross-terms.

To forge ahead analytically, we first assume that each of the two phases in our toy IGM also have different densities, each uniform. We indicate the average density of bubble material as $\langle \delta \rangle_i$, with the density in the bulk IGM $\langle \delta \rangle_n$, enforced by continuity such that
\begin{equation}
    \langle \delta \rangle_n = - \langle \delta \rangle_i \frac{Q}{1-Q} .
\end{equation}
The only unknown here is $\langle \delta \rangle_i$. One could treat it as a free parameter, or parameterize it flexibly as a function of redshift and/or bubble size. We discuss this possibility further in \S\ref{sec:bubble_density}, where we introduce a simple model for $\langle \delta \rangle_i$.

Now, because we are assuming that neutral and ionized patches are of uniform (but redshift-dependent) densities, we can use a halo-model-like argument to write down the two-point terms involving $b$ and $\delta$. For example, $\langle b \delta^{\prime} \rangle$ will be the sum of two terms: one in which the primed point is neutral, and one in which it is ionized, in each case weighted by the density of the relevant medium:
\begin{equation}
    \langle b \delta^{\prime} \rangle =  \langle \delta \rangle_i \prob_{bb} + \langle \delta \rangle_n \prob_{bn} \label{eq:P_bd}
\end{equation}
where we have used the previous notation to indicate the probability that two points are in bubbles, $\prob_{bb}$, and a new term $\prob_{bn}$ to indicate the probability that only one point is in a bubble. We can write the latter as
\begin{align}
    \prob_{bn} & = (1 - \prob_1) \exp\bigg[- \int dR_b n_b(R_b) \left(V(R_b) - V_o(r, R_b) \right) \bigg] \nonumber \\
    & \times \left\{1 - \exp\bigg[- \int dR_b n_b(R_b) \left(V(R_b) - V_o(r, R_b) \right) \bigg] \right\}  .
\end{align}
In words, the above equation computes the probability that a single source can ionize one point but \textit{not} the other (term in curly brackets), times the probability that the other point is \textit{not} ionized by a different source that leaves the first point untouched (first exponential term), times the probability that a single source does not reside in the overlap volume and engulf both points (leading $1-\prob_1$ factor).

From Eq. \ref{eq:P_bd}, it is clear that if the density of ionized material is order unity ($\langle\delta\rangle_i\sim1$), the cross-term $\langle b \delta^{\prime} \rangle$ will be comparable to the auto term $\langle b b^{\prime} \rangle$, so long as $Q \ll 1$  and $P_{bn} \ll P_{bb}$. The first condition is plausible, and we will find that indeed $\langle \delta \rangle_i \approx 1$ when $Q \lesssim 0.2$. The second condition is less clear; certainly, on small scales, the $P_{bn}$ term should be suppressed significantly by the $1 - P_1$ factor, since two points are increasingly likely to be in the same bubble on progressively smaller scales. However, on large scales, $P_{bn} \rightarrow Q (1 - Q)$, which means $P_{bn} \geq P_{bb}$ when $Q \leq 0.5$.

This is a curious feature of this model: for $\langle b \delta^{\prime} \rangle \simeq \langle b b^{\prime}\rangle$, it is possible that $\xi_{21}$ becomes negative. Just comparing 2-pt terms, $\langle b \delta^{\prime} \rangle = 2 \langle b b^{\prime} \rangle$, this will occur when
\begin{equation}
    \frac{P_{bn}}{P_{bb}} < 2 \left(\frac{2 \langle \delta \rangle_i - 1}{\langle \delta \rangle_i}\right) \left( \frac{1-Q}{Q} \right) \label{eq:P_bn_condition} .
\end{equation}
Furthermore, if $\langle \delta \rangle_i < 0.5$, the RHS of Eq. \ref{eq:P_bn_condition} becomes negative, a condition that cannot be satisfied since $P_{bn}$ and $P_{bb}$ are both positive.

Such strong contributions from cross-terms involving ionization and density are not expected from more physically motivated models, but it is not surprising that they can become significant in our framework given the assumption of sharp, spherical bubbles and a perfect two-zone IGM. However, in detail, the amplitude and sign of $\xi_{21}$ depend not only on the 2-pt contributions, but also on higher order terms, which we discuss next.

\subsubsection{3- and 4-pt contributions to the power spectrum} \label{sec:crossterms_Npt}
The two most obvious ways to proceed with the remaining terms in Eq. \ref{eq:xi21_full} are to (i) neglect them, or (ii) fully embrace the binary framework and write down these terms following the logic applied to the $\langle b \delta^{\prime} \rangle$ term above. Though higher order terms are likely to be smaller than the two-point terms, at least on scales $k \lesssim 1 \ \invMpch$, they are not negligible in general \citep[e.g.,][]{Lidz2007,Georgiev2021}. Both options have some undesirable properties.

For example, option (i) must artificially set $\langle b \delta \rangle = 0$ in order to ensure that fluctuations vanish on large scales, despite the fact that correlations between bubbles and density imply that $\langle b \delta \rangle \neq 0$. However, the three and four-point terms $\langle b \delta \delta^{\prime} \rangle$  and $\langle b b^{\prime} \delta \delta^{\prime} \rangle$ can \textit{not} be set to zero, otherwise 21-cm fluctuations will not vanish as $Q \rightarrow 1$ as they must. This is apparent from Eq. \ref{eq:xi21_full} -- the leading factor of the matter fluctuations, $\langle \delta \delta^{\prime} \rangle$, will persist regardless of $Q$, and so non-zero contributions from other terms involving the fluctuation $\delta \delta^{\prime}$ are required in order for 21-cm fluctuations to vanish as $Q \rightarrow 1$. More on this momentarily.

Regarding option (ii), the binary field model predicts:
\begin{align}
    \langle b b^{\prime} \delta \rangle & = \langle \delta \rangle_i \prob_{bb} \\
    \langle b \delta \delta^{\prime} \rangle & = \langle \delta \rangle_i^2 \prob_{bb} + \langle \delta \rangle_i \langle \delta \rangle_n \prob_{bn} \\
    \langle b b^{\prime} \delta \delta^{\prime} \rangle & = \langle \delta \rangle_i^2 \prob_{bb} \label{eq:bbdd}
\end{align}
A few observations about these terms:
\begin{itemize}
    \item The only additional $k$-dependent suppression of 21-cm power beyond that caused by the two-point term $\langle b \delta^{\prime} \rangle$ comes from the $\langle b \delta \delta^{\prime} \rangle$ term above, which has a single leading factor of $\alpha=-1$ (see Eq. \ref{eq:xi21_full}), in contrast to the $\langle b b^{\prime} \delta \rangle$ and four-point terms which are both positive.
    \item If we want to reduce the contribution from these higher order terms by, e.g., setting \ref{eq:bbdd} to zero, we can only do so if we also set $\langle b \delta \rangle = 0$. Otherwise, fluctuations will not vanish on large scales (see last two terms of Eq. \ref{eq:xi21_full}). However, this on its own will violate the requirement that fluctuations vanish as $Q \rightarrow 1$ (see above).
    \item On large scales, the leading factor of $\prob_{bb}$ means that the contribution of higher order terms will grow as reionization proceeds ($\prob_{bb} \rightarrow Q^2$).
    \item The power spectrum of a binary density field that traces the bubble field will exhibit a sharp feature on the typical bubble scale and no structure on smaller scales, at odds with the well-understood shape of the matter power spectrum.
\end{itemize}
Given these challenges, we employ a third option, which ensures that 21-cm fluctuations vanish on large scales and as $Q \rightarrow 1$. From Eq. \ref{eq:xi21_full}, it is clear that the latter condition requires
\begin{equation}
    -2 \alpha \langle b \delta \delta^{\prime} \rangle - \alpha^2 \langle b b^{\prime} \delta \delta^{\prime} \rangle \rightarrow \langle \delta \delta^{\prime} \rangle
\end{equation}
as $Q \rightarrow 1$. We take
\begin{align}
    \langle b \delta \delta^{\prime} \rangle & = Q \langle \delta \delta^{\prime} \rangle \\
    \langle b b^{\prime} \delta \delta^{\prime} \rangle & = \langle b b^{\prime} \rangle \langle \delta \delta^{\prime} \rangle + \langle b \delta^{\prime} \rangle^2 + \langle b \delta \rangle^2 .
\end{align}
The second expression invokes Wick's theorem, common in the literature despite expectations that the 21-cm field is non-Gaussian, while the first is the simplest treatment that recovers the desired limiting behaviour. Together, these terms can be thought of as a correction factor applied to the matter fluctuations in Eq. \ref{eq:xi21_full}, $\langle \delta \delta^{\prime} \rangle \rightarrow (1-Q)^2 \langle \delta \delta^{\prime} \rangle$.

Finally, we set $\langle bb^{\prime} \delta = 0$ to satisfy the requirement that $\xi_{21} \rightarrow 0$ on large scales.


\subsubsection{Model for the density of bubble material} \label{sec:bubble_density}
Critical to the simple cross-term treatment described above is knowledge of the density of bubble material, $\langle \delta \rangle_i$. To determine this bubble density, we make an argument similar to abundance matching in galaxy formation models: we assume that if a fraction $Q$ of the IGM is in bubbles, then that volume is also the densest fraction $Q$ of the IGM. Then, our task is to determine the density threshold above which $Q$ per-cent of the IGM resides.


To do this, we first define the variance of fluctuations at redshift $z$ over a region of radius $R$ to be
\begin{equation}
    \sigma_R^2 = \int \dfrac{d^3 k}{(2\pi)^3} P(k,z) |W_R(k)^2|,
\end{equation}
where $P(k,z)$ is the matter power spectrum at $z$, and $W_R(k)$ is a window function encoding the shape of the region. We assume a spherical top-hat, which has the form
\begin{equation}
    W_R(k) = \dfrac{3}{(kR)^3}\left[\sin(kR)-(k R) \cos(k R)\right] .
\end{equation}


We further assume that the PDF of the density field, $\mathcal P(\delta_R)$, is log-normal \citep{Coles1991,Bi1997}.
Then, if a fraction $Q$ of the volume of the universe is ionized, we can associate that with a minimum density $\delta_R^{\rm min}$ through
\begin{equation}
\int_{\delta_R^{\rm min}}^\infty d\delta_R \mathcal P(\delta_R) = \dfrac{1}{2} {\rm erfc}\left(\dfrac{\delta_R^{\rm min}}{\sqrt{2} \sigma_R}\right) = Q,
\end{equation}
or equivalently
\begin{equation}
    \delta_R^{\rm min}(Q) = \sqrt{2} \sigma_R \, {\rm erfc}^{-1}(2Q).
\end{equation}
Now it is easy to compute the density of ionized material,
\begin{equation}
\langle \delta \rangle_i = \int_{\delta_R^{\rm min}}^\infty d\delta_R \mathcal P(\delta_R) \delta_R = \exp[-(\delta_R^{\rm min})^2/(2\sigma_R^2)] \dfrac{\sigma_R}{\sqrt{2\pi}}.
\end{equation}
The only free parameter of this model is the smoothing scale $R$ employed to compute the variance in the density field, which we will hereafter refer to as $R_{\rm{sm}}$. There are only a few natural length scales in our model thus far, all of which are related to the characteristic bubble size. We will explore several possibilities in \S\ref{sec:21cmfast_mocks}, and compare to the mean density of ionized gas and ionization -- density cross spectrum from \textsc{21cmfast} models for guidance.

Note that in this section, we effectively \textit{have} made a physical argument about the nature of reionization, namely, that it occurs ``inside out.'' Though this choice departs from our effort to avoid explicit astrophysical assumptions, it is appropriate for comparisons to \textsc{21cmfast}, and could be generalized in the future \citep[see, e.g.,][for one approach]{Pagano2020}. Fortunately, scenarios with a strong degree of \textit{anti-correlation} between ionization and density fields generate stronger 21-cm fluctuations than the alternative, and should thus be easier to rule out as upper limits become more stringent \citep{Pagano2020,Pagano2021}.

\begin{figure*}
\begin{center}
\includegraphics[width=0.98\textwidth]{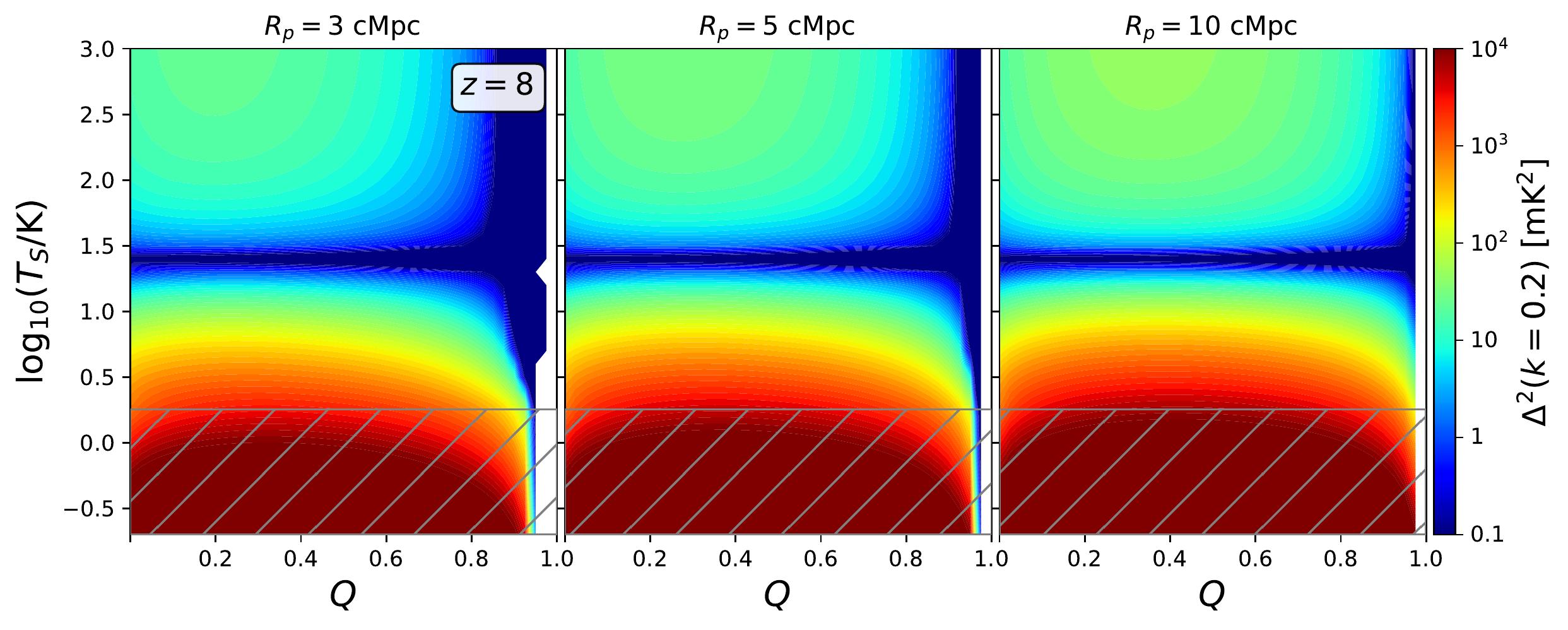}
    \caption{{\bf 2-D parameter study showing the effects of ionized fraction $Q$ and spin temperature $\TS$ for log-normal bubble size distributions.} The color-scale indicates the dimensionless power, $\Delta_{21}^2$, at $z=8$ and  $k=0.2 \ \invMpch$. We assume $\sigma_b=1$ for characteristic bubble sizes of 1, 5, and 10 cMpc (left to right). The horizontal band in each panel corresponds to $\Ts=\Tcmb$ at $z=8$, where the brightness temperature goes to zero, while the cross-hatched region in the bottom indicates spin temperatures below the adiabatic limit for a uniform medium at $z=8$, $T_{\rm{adi}}=1.82$ K.
    }
\label{fig:grid2d}
\end{center}
\end{figure*}

\subsection{Redshift space distortions} \label{sec:rsds}

We now add the effect of peculiar velocities \citep[see also, e.g.,][]{Kaiser1987,Barkana2005RSD}.
These give rise to redshift-space distortions (RSDs), which to linear order we can write as
\begin{equation}
  \delta T_b = T_0(z) (1 + \alpha b) (1 + \delta - \delta_v) (1 + \delta_{\varphi}), \label{eq:dTb_bubbles_RSD}
\end{equation}
where $\delta_v$ is the line-of-sight velocity-gradient anisotropy.
This last quantity is given (again to linear order and during the epoch of interest) by $\delta_{v} = -\mu^2 \delta$ in Fourier space, where $\mu = k_{||}/k$,
so we have
\begin{equation}
  \delta T_b = T_0(z) (1 + \alpha b) [1 + (1 + \mu^2 + \beta_{\varphi})\delta ]. \label{eq:dTb_bubbles_mu}
\end{equation}
We ignore non-linear RSDs \citep[e.g.,][]{Mao2012,Jensen2013} -- see~\citet{Greig:2018hja} for the implementation in \textsc{21cmfast}. We also neglect the light cone effect \citep[see, e.g.,][for detailed treatments]{Datta2012,LaPlante2014,Chapman2019}. This is a conservative approach given that these effects boost the power spectrum, and so will drive lower limits on the IGM spin temperature upward once included.

In traditional galaxy surveys one can measure different modes $\vec k$, and thus probe the $\mu$ dependence of the RSDs.
In 21-cm studies, however, the situation is different.
Foregrounds impose an observational cutoff, as small wavenumbers along the line of sight are inaccessible for cosmology \citep[e.g.,][]{Datta2010,Parsons2012a,Morales2012,Vedantham2012,Thyagarajan2013,Liu2014a}.
This ``foreground wedge'' in fact occupies the majority of the Fourier plane, so in practice the modes observed at any fixed spherical $k$ are chiefly along the line of sight, with $\mu \approx 1$\footnote{For HERA $\mu \gtrsim 0.97$, with values closer to one indicative of increasingly aggressive wedge cuts.}.
We will therefore often simply fix $\mu=1$.
Alternatively, when comparing to \textsc{21cmfast} simulations we will manually set $\mu=0.6$, which recovers $\langle(1+\mu^2)^2\rangle=1.87$, the (linear) spherically-averaged value of RSDs. Note that the use of a single average value of $\mu$ is likely to be overly simplistic due to the non-linear relationship between density and ionization \citep{Pober2015wedge}. We defer a more detailed treatment to future work.


Upon including RSDs, we take $\delta \rightarrow \delta - \delta_v$, resulting in the following modifications to cross-terms:
\begin{align}
    \langle b \delta^{\prime} \rangle & \rightarrow \langle b \delta^{\prime} \rangle - \langle b \delta_v^{\prime} \rangle \\
    \langle b \delta \delta^{\prime} \rangle & \rightarrow  \langle b \delta \delta^{\prime} \rangle -  \langle b \delta \delta_v^{\prime} \rangle -  \langle b \delta^{\prime} \delta_v \rangle +  \langle b \delta_v \delta_v^{\prime} \rangle \\
    \langle b b^{\prime} \delta \delta^{\prime} \rangle & \rightarrow  \langle b b^{\prime} \delta \delta^{\prime} \rangle -  \langle b b^{\prime} \delta \delta_v^{\prime} \rangle -  \langle b b^{\prime} \delta^{\prime} \delta_v \rangle +  \langle b b^{\prime} \delta_v \delta_v^{\prime} \rangle
\end{align}
where we have shown for completeness the 3- and 4-pt terms, despite neglecting them in what follows.



\subsection{Density fluctuations} \label{sec:dd}
Finally, $\langle \delta \delta^{\prime} \rangle$ is equivalent to the matter correlation function, $\xi_{\delta \delta}$ -- we compute the linear matter power spectrum using \textsc{camb} \citep{Lewis2000} and Fourier transform to obtain $\xi_{\delta \delta}$, i.e.,
\begin{equation}
    P_{\delta \delta}(k) = \frac{1}{(2\pi)^3} \int 4\pi R^2  \frac{\sin(kR)}{kR} \xi_{\delta \delta}(R) dR \label{eq:hankel_transform}
\end{equation}
which is the same operation used to convert $\xi_{21}$ to the 21-cm power spectrum in Eq. \ref{eq:P21}. The highly-oscillatory nature of these integrals pose a challenge -- we include options for Clenshaw-Curtis integration \citep[as in, e.g.,][]{Diemer2018} and FFTLog algorithms \citep{Talman78,Hamilton00}, which are implemented in \textsc{scipy} and \textsc{mcfit}\footnote{\url{https://github.com/eelregit/mcfit}}, respectively. The \textsc{mcfit} approach is generally faster by $\sim 2-3$x, and so is the default in \textsc{micro21cm}.

\subsection{Putting it all together}
In Figure \ref{fig:components}, we assemble a representative 21-cm power spectrum term by term. In the left panel, we start in a $z=8$ IGM with uniform temperature $\TS=1.8$ K, and only matter fluctuations (dotted black). In reality, gas density and kinetic temperature are coupled due to adiabatic cooling -- the blue dashed curve accounts for this correlation, which suppresses power on all scales since denser regions are also warmer than less dense regions (see \S\ref{sec:adiabatic_corr}). These first two cases ignore redshift-space distortions (see \S\ref{sec:rsds}). The dot-dashed cyan curve averages the power spectrum over all $\mu$, providing a boost in power that nearly cancels out the suppression caused by the $\contrast$ correction. Finally, we note that many current experiments almost exclusively probe line of sight modes, with $\mu \sim 1$, which we show in the solid magenta curve \citep[see also Fig. 3 in][]{HERA2021Theory}. The ratio of each case to the matter-only case is shown in the bottom left panel.

Next, in the right panel of Fig. \ref{fig:components}, we start from the dot-dashed cyan curve of the left panel and add ionized bubbles with a log-normal size distribution (see \S\ref{sec:bubbles}). We further assume an ionized fraction $Q=0.4$, and a typical bubble size of $R_b=5$ cMpc, which are reasonable choices for $z \sim 8$. If reionization were spatially homogeneous, the power would be suppressed by a factor of $(1-Q)^2$ at all $k$, which we show in the opaque cyan curve. However, a model with discrete bubbles boosts power around the bubble scale, as seen in the orange-dotted curve, which adds the ionization auto-correlation term only (see Eq. \ref{eq:xi21_full}). Finally, the red curve adds the cross-terms involving both ionization and density following the procedure of \S\ref{sec:crossterms_2pt} and \ref{sec:crossterms_Npt}. In the bottom row, we once again show the ratio between each power spectrum in the top panel with the matter-fluctuations-only case (dotted black).

Given that many current experiments probe large $k \simeq 0.2 \invMpch$ scales, largely at frequencies $\nu \gtrsim 100$ MHz \citep[e.g.][]{Barry2019,Mertens2020,HERA2021} where the bulk of reionization is expected to occur \citep[e.g.,][]{Robertson2015,Bouwens2015,Mason2015,Gorce2018,Finkelstein2019}, in Fig. \ref{fig:grid2d} we show predictions for the large-scale $k = 0.2 \ \invMpch$ power over all of $(Q, \Ts)$ space. From left-to-right we increase the typical bubble size from 3, to 5, to 10 cMpc, and color-code by $\Delta^2(k=0.2)$ from $1$ to $10^4 \ \mathrm{mK}^2$ (blue to red colours). The cross-hatched region in the bottom of each panel indicates temperatures below $1.8$ K, which is the expected minimum temperature of an adiabatically-cooled $z=8$ IGM, making it clear that $10^4 \ \rm{mK}^2$ signals (red) require a super-cooled IGM.

Focusing next on the $10^3 \ \rm{mK}^2$ range (orange), which is comparable to recent limits, we see that viable scenarios generally require $\TS \simeq 3-10 \ \rm{K}$. If bubbles are small (left panel), sub-adiabatic temperatures may be required, but for larger bubbles, $R_p = 5$ or 10 cMpc (center, right columns), $10^3 \ \rm{mK}^2$ fluctuations are possible without sub-adiabatic temperatures provided that reionization is not just beginning or just ending. Ionization fluctuations are maximized near the reionization midpoint, in which case, such fluctuations can be achieved if the IGM temperature is $\log_{10}(\Ts /{\rm{K}}) \simeq 0.5$, i.e., if $\Ts \simeq 3$ K, consistent with the interpretation of HERA's recent $10^3 \ \rm{mK}^2$ limits \citet{HERA2021Theory}.

In general, as the strength of ionization fluctuations grow, 21-cm fluctuations will also grow stronger if the temperature is held fixed. As ionization fluctuations decline in the latter half of reionization, holding the power constant demands that the spin temperature deviate more strongly from $\TR$. There are two exceptions to this behaviour. At early times, if bubbles are small (left column), the assumed positive correlation between ionization and density results in a decline in the power as $Q$ grows (at fixed $\Ts$). Second, if the spatial scale of interest is much larger than the typical bubble size, $k \lesssim R_p^{-1}$, the amplitude of fluctuations on that scale depend very little on the ionized fraction.

Figure \ref{fig:grid2d} provides a means of rough, by-eye inference. Provided our phenomenological model is reasonably accurate, one can simply ``read off'' the $Q$, $\Ts$, and $R$ values that are consistent with new power spectrum upper limits. Though we have some indication already that the phenomenological model performs well compared to more sophisticated calculations, e.g., the association of $10^3 \ \rm{mK}^2$ fluctuations at $k=0.2 \ \invMpch$ with $\Ts \simeq 3$ K \citep[see above; also][]{HERA2021Theory}, in the next section, we provide a much more detailed comparison to \textsc{21cmfast} calculations.

\section{Comparison to \textsc{21cmfast} Models} \label{sec:21cmfast_mocks}
Having outlined the various components of our phenomenological model, we now compare its predictions to two illustrative models generated with \textsc{21cmfast} \citep{Mesinger2007,Mesinger2011,Murray2020}. Our goal is to assess the accuracy of the model relative to more sophisticated calculations, test various modeling choices, and set expectations for interpreting fits to \textsc{21cmfast} mocks in \S\ref{sec:fits}.

\begin{table}
\begin{tabular}{| l | l | c | c | }
\hline
ID & model name & $\Mturn$ & $\log_{10}(L_X / \mathrm{SFR})$  \\
\hline
1a & slow / no heat & n/a & 37.5 \\
1b & slow / cold & n/a & 38.5 \\
1c & slow / warm & n/a & 39.5 \\
\bf 1d & \bf slow / hot (EOS21) & \bf n/a & \bf 40.5 \\
2a & fast / no heat & $10^9\,M_\odot$ & 37.5 \\
2b & fast / cold & $10^9\,M_\odot$ & 38.5 \\
2c & fast / warm & $10^9\,M_\odot$ & 39.5 \\
2d & fast / hot & $10^9\,M_\odot$ & 40.5 \\
\hline
\end{tabular}
\caption{{\bf Summary of \textsc{21cmfast} models.}
We use two reionization models: slow (corresponding to the EOS21 parameters of \citet{MunozEoS}) and fast (which have the same PopII parameters but a nonzero turnover mass $M_{\rm turn}$, and thus no PopIII stars).
For each of those two models we vary the X-ray heating efficiency parameter $L_X$ as indicated in this table.
For the entire set of galaxy parameters, and how they fit all current EoR data, see \citet{MunozEoS}.}
\label{tab:21cmfast}
\end{table}

We will compare to two benchmark \textsc{21cmfast} models.
The first, model \#1, has the same set of parameters as the {\tt AllGalaxies} simulations of \citet{MunozEoS}, and thus includes atomic-cooling galaxies forming Pop~II stars \citep[following][]{Park2019} as well as molecular-cooling halos forming Pop~III \citep[following][]{Qin2020}, with joint feedback from Lyman-Werner photons \citep{Haiman1997,Machacek2001,Visbal2014} and streaming velocities \citep{Tseliakhovich2010,Visbal2012,MunozVAOs}.
The second model, on the other hand, imposes a cutoff for star formation at $M_{\rm turn}=10^9\,M_\odot$, so halos below that mass do not form stars.
As a consequence, there are no Pop~III stars in that model, and the evolution of the 21-cm signal is faster.
The rest of galaxy properties are the same between the two models, with star formation parameters calibrated to high-$z$ luminosity functions \citep[from][]{Finkelstein2015} and X-ray spectra representative of X-ray binaries hardened by neutral columns expected of low-mass galaxies at high-$z$ \citep{Das2017}. The parameter values for each \textsc{21cmfast} model are summarized in Table \ref{tab:21cmfast}.

The only parameter we will vary is the X-ray luminosity of the first galaxies, as it strongly affects the values of $T_S$ during the epoch of interest.
We will start with the fiducial choice  of $\log_{10}(L_X / \mathrm{SFR})=40.5$, which is $\sim 10$x higher than that generated by high-mass X-ray binaries in nearby star-forming galaxies \citep{Mineo2012}, as expected of low-metallicity environments at high redshift \citep[e.g.][]{Fragos2013,Brorby2016}. We explore this parameter in order-of-magnitude steps down to 37.5, so as to cover a broad range of possibilities.

\begin{figure}
\begin{center}
\includegraphics[width=0.45\textwidth]{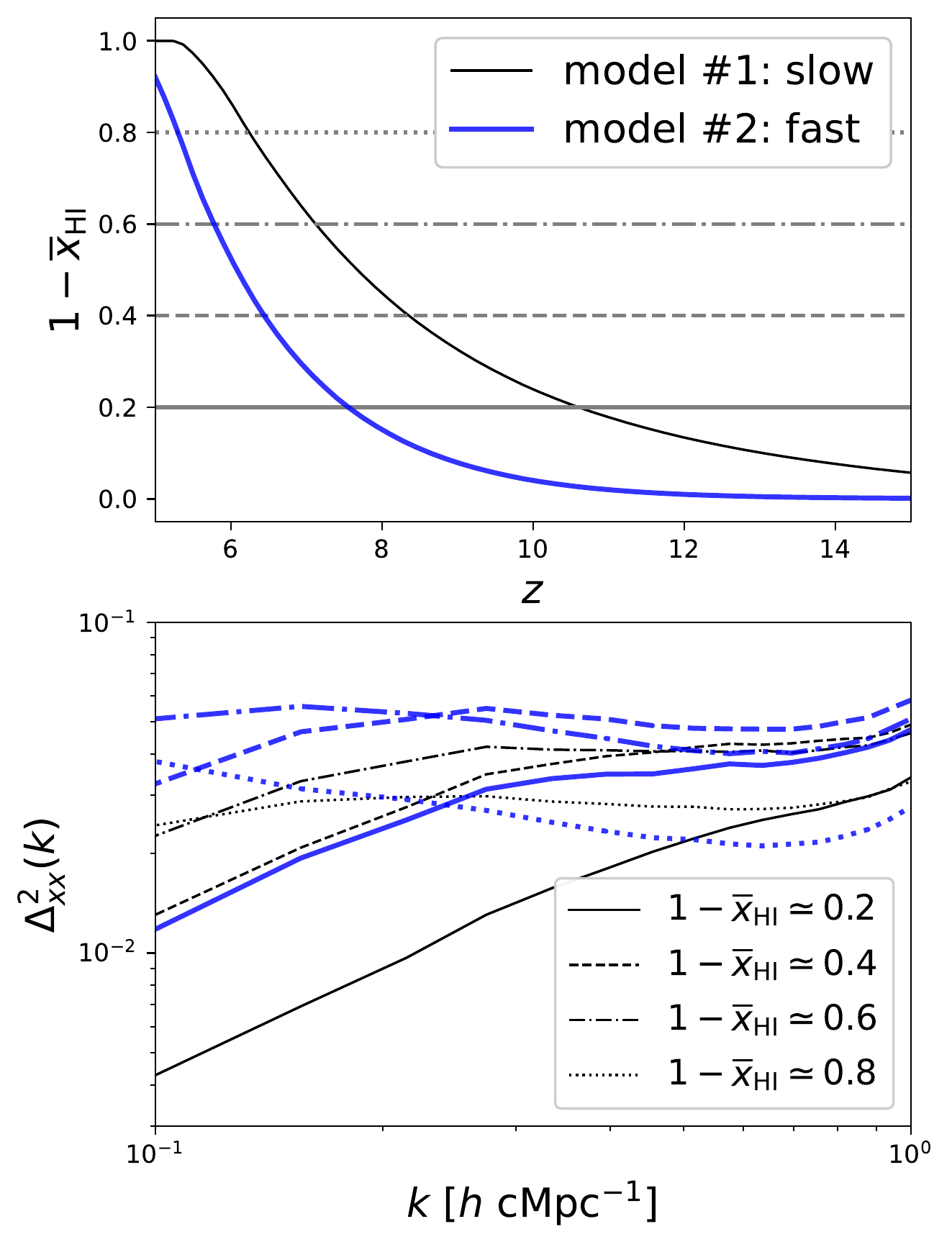}
    \caption{{\bf Example \textsc{21cmfast} models used for comparison throughout.} Mean ionization histories (top) and ionization power spectra (bottom) at four different mean ionized fractions, $Q$. Both models are `late reionization' scenarios, differing largely in the duration of reionization, with the ``slow'' model (black), the transition from 20\% to 80\% ionization takes $\Delta z \simeq 4$, while in the ``fast'' scenario (blue), reionization begins later and the 20\% to 80\% duration is $\Delta z \simeq 2$. The ``fast'' scenario generates stronger fluctuations at fixed ionized fraction, as the sources of reionization are more biased.}
\label{fig:mocks_21cmfast}
\end{center}
\end{figure}

\begin{figure*}
\begin{center}
\includegraphics[width=0.98\textwidth]{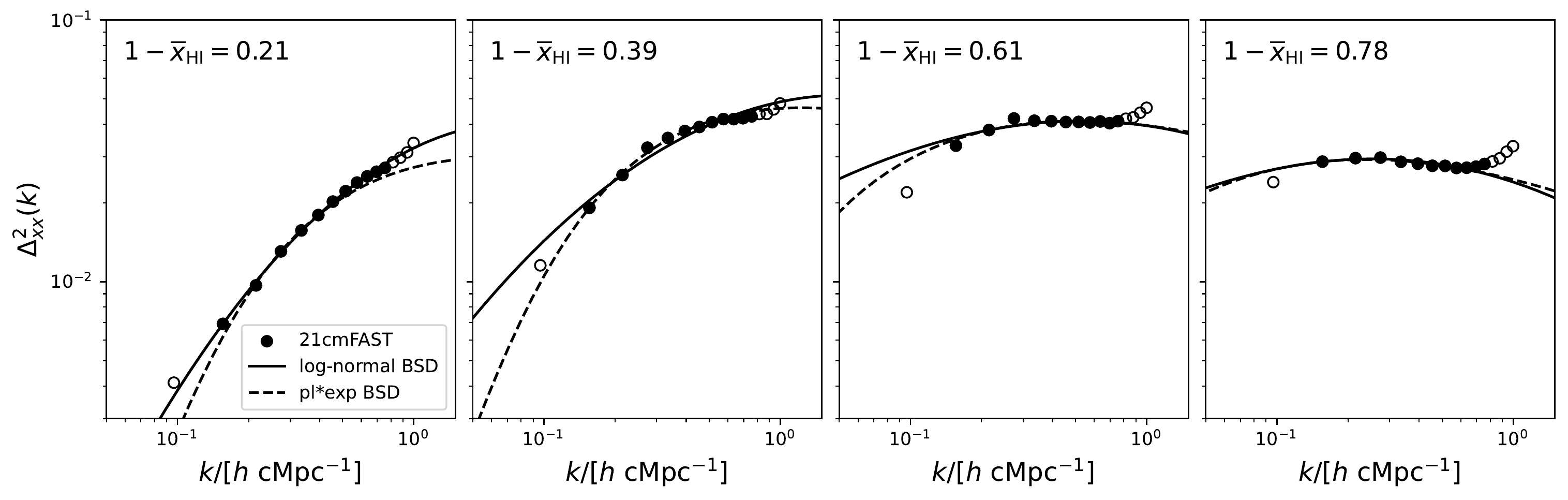}
\caption{{\bf Comparison of phenomenological model (lines) to \textsc{21cmfast} ionization power spectra (points)} for the ``slow'' reionization scenario. From left to right, we compare at fixed mean ionized fraction from $Q\simeq0.2$ to $Q\simeq0.8$, as indicated in the upper left corner of each panel, assuming that $Q = 1 - \xHIavg$. We fit the \textsc{21cmfast} points to calibrate $R$ and $\sigma$ (or $\gamma$) for a fair comparison. Dashed lines correspond to the power-law-times-exponential BSD, while solid lines correspond to the log-normal BSD.
}
\label{fig:Pxx_v_21cmfast_cal}
\end{center}
\end{figure*}

We show the mean ionization histories (top) and ionization power spectra (bottom) for both our models in Fig.~\ref{fig:mocks_21cmfast}. Reionization occurs more gradually in model \#1 than in model \#2, so we dub them ``slow'' and ``fast,'' respectively, though they are both `late reionization' scenarios, with neutral fractions of $\sim 20$\% at $z \sim 6$, in accordance with recent constraints \citep{Becker2015,Bosman2021,Keating2020,Qin2021}. We will examine all four possibilities for the spin temperature evolution in \S\ref{sec:21cmfast_fits} but for now focus only on the ionization field.
In Fig. \ref{fig:Pxx_v_21cmfast_cal} we compare the ionization power spectrum computed by our model to those from \textsc{21cmfast} at the same mean ionized fraction, $1-\xHIavg$, or equivalent volume filling fraction, $Q$ in the phenomenological model\footnote{Note that this is itself a potential source of uncertainty in our comparison, since the ionization field is not binary in \textsc{21cmfast}.}. At a series of ionized fractions spanning from 20\% to 80\%, we show power spectra drawn from \textsc{21cmfast} model 1 (points), and a best-fit representation using the phenomenological model. The parameters of a log-normal BSD ($R$ and $\sigma$; solid lines) and power-law-with-exponential-cut-off model ($R$ and $\gamma$; dashed lines) are calibrated to match the \textsc{21cmfast} power spectra at $0.1 \leq k /(\invMpch) \leq 0.8$. Overall, the shape of the ionization power spectrum can be well-modeled using either BSD parameterization. We exclude points at $k > 0.8 \invMpch$ from the fit, as discrete sampling effects start to become apparent at smaller scales and so would artificially bias the calibration.

\begin{figure}
\begin{center}
\includegraphics[width=0.49\textwidth]{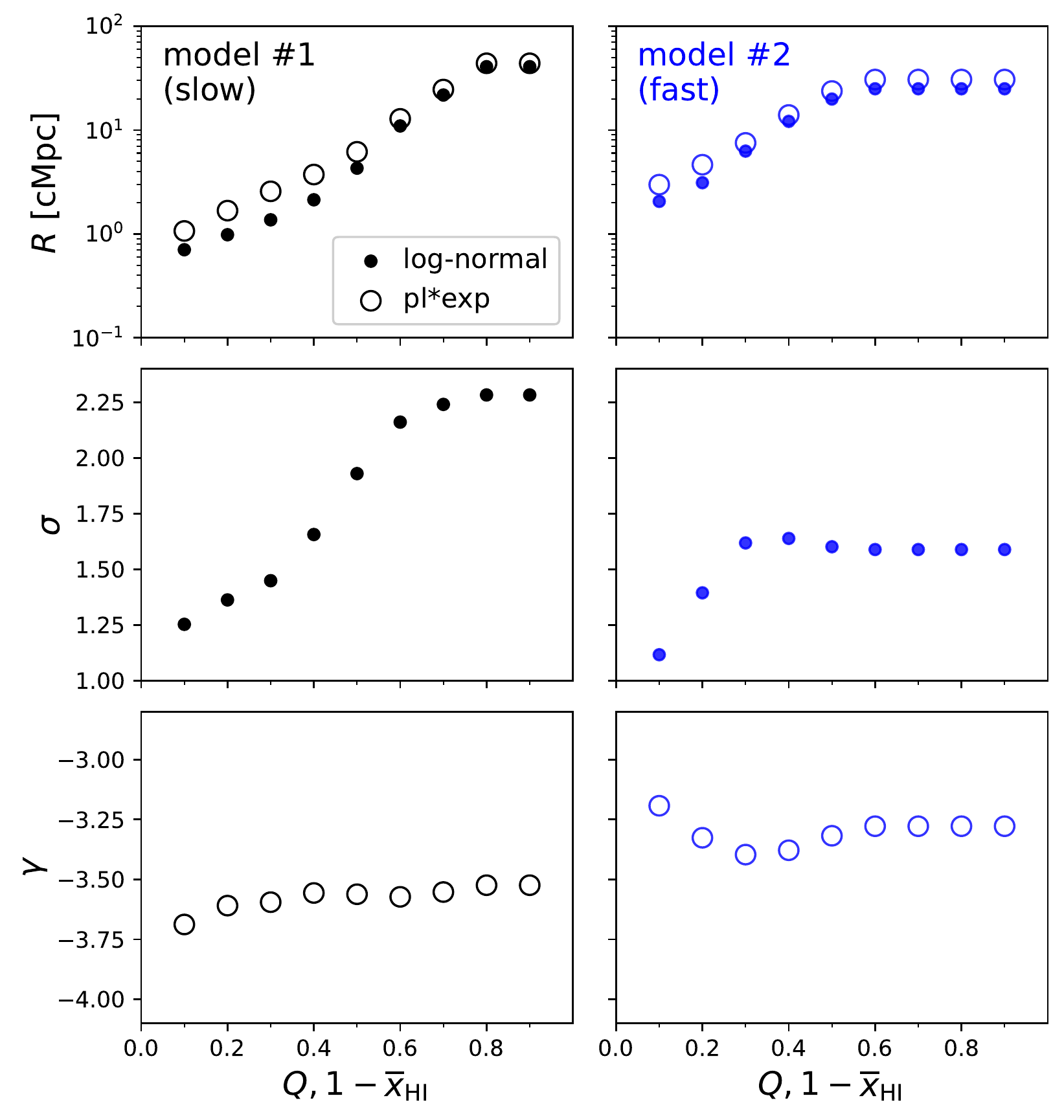}
\caption{{\bf Evolution of BSD parameters inferred from fits to example \textsc{21cmfast} ionization power spectra.} The typical bubble size evolution is well-captured by a power-law dependence on $Q$ (top) for both ``slow'' and ``fast'' scenarios (left and right columns, respectively), while $\sigma$ exhibits a gradual rise and eventual plateau (middle row). The power-law slope $\gamma$ is relatively constant for both reionization scenarios (bottom row).}
\label{fig:Pxx_v_21cmfast_evol}
\end{center}
\end{figure}

In Fig. \ref{fig:Pxx_v_21cmfast_evol}, we show the best-fit BSD parameters as a function of $Q$ for both \textsc{21cmfast} scenarios (left and right columns). For both BSD models, the power spectra are consistent with rapidly growing bubble sizes (top row). For the log-normal BSD, the dispersion $\sigma$ gradually increases from $\sigma \simeq 1.25$ to 2.25 (second row). The power-law-times-exponential BSD varies less with $Q$; generally $\gamma$ lies between $\gamma \sim -3.75$ and $\gamma \sim -3.25$.

With calibrated $R(Q)$ and $\sigma(Q)$ values in hand, we now compare the phenomenological model's prediction for the density of bubble gas to the results extracted from \textsc{21cmfast} boxes. As discussed in \S\ref{sec:bubble_density}, the key choice in the phenomenological model is the scale on which to smooth the density field when computing its variance, which sets the fraction of the volume above a given density contour. Here, we explore three options, which assume a smoothing scale equal to the scale of (i) the peak of the volume-weighted, logarithmic BSD, $V dn/d\log R$, (ii) the peak of the volume-weighted, linear BSD, $V dn/dR$, and (iii) the radius at which the joint probability of ionization, $\langle b b^{\prime} \rangle$ is no longer equivalent to the one-bubble term. The final option requires choosing a threshold, e.g., the scale at which $P_1 = X \langle b b^{\prime} \rangle$, with $X$ a free parameter. It is not obvious how to choose this critical threshold, or if it is more or less meaningful than options (i) or (ii) -- it merely serves as another approach to employ in comparisons with \textsc{21cmfast}. We show cases for a threshold of $0.97 \pm 0.025$, which result in densities that generally lie between the predictions of options (i) and (ii) described above.

\begin{figure}
\begin{center}
\includegraphics[width=0.45\textwidth]{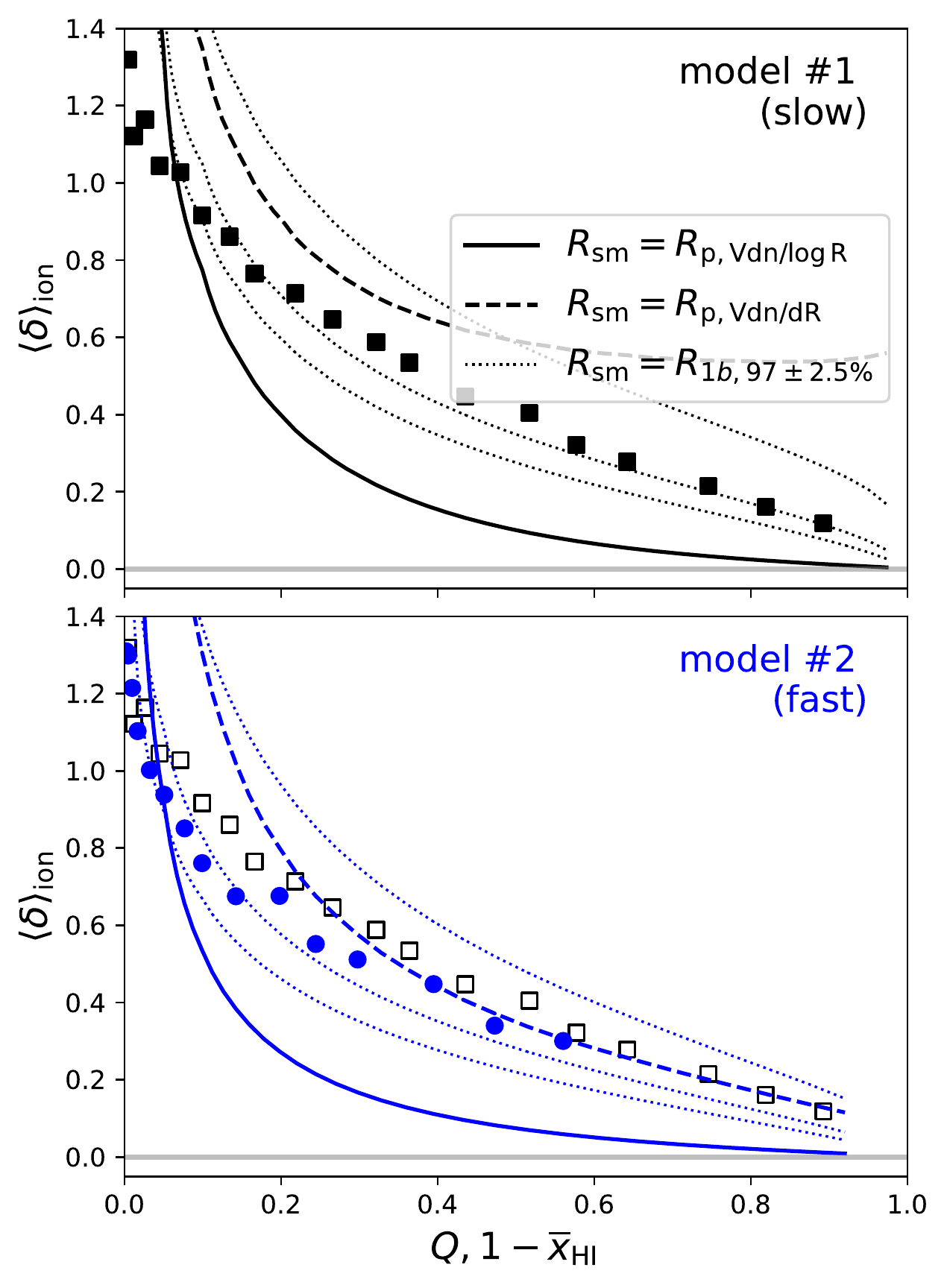}
    \caption{{\bf Mean density of ionized gas in \textsc{21cmfast} (points) and phenomenological models (lines) vs. ionized fraction} for the ``slow'' (top) and ``fast'' EoR scenarios  (bottom). We show predictions for several different smoothing scales (see \S\ref{sec:bubble_density}), which sets the variance in the density field and thus density contour containing $Q$\% of the volume, as indicated in the legend in the top panel. We have assumed the log-normal BSD, with $R(z)$ and $\sigma(z)$ calibrations shown in Figs. \ref{fig:Pxx_v_21cmfast_cal} and  \ref{fig:Pxx_v_21cmfast_evol}. Black points are repeated in both panels for ease of comparison.
    }
\label{fig:bubble_density}
\end{center}
\end{figure}

In Fig. \ref{fig:bubble_density}, we show the bubble density predictions for the log-normal BSD model compared to \textsc{21cmfast}. While each approach results in the correct behaviour qualitatively, none provide an accurate match at all $Q$ or in both reionization scenarios. In general, the density of ionized gas evolves much more rapidly in the phenomenological model than in \textsc{21cmfast} models suggest at early times, $Q \lesssim 0.3$. While model \#2 is well-matched by smoothing scale option (ii) described above (dashed curves; bottom panel), at least at $Q \gtrsim 0.2$, the same approach does not provide as good a match for model \#1 (dashed curve; top panel). For a suitably chosen threshold, a smoothing scale linked to the decline of the one-bubble term does result in slightly more gradual evolution (dotted curves). Though the solid curves do not provide as good of a match in general, we adopt them in all that follows for reasons we discuss further momentarily.

\begin{figure*}
\begin{center}
\includegraphics[width=0.98\textwidth]{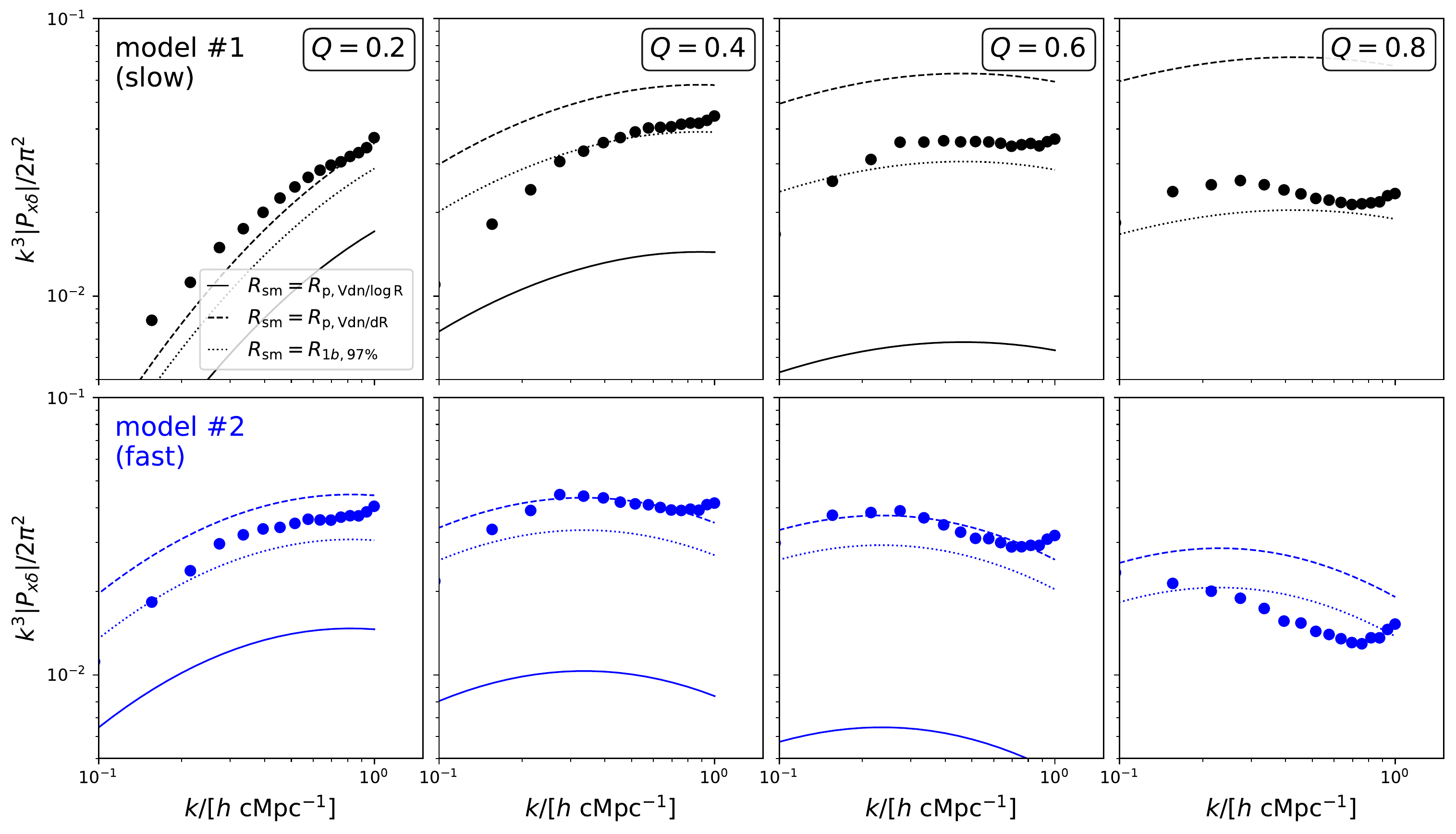}
    \caption{{\bf Cross spectrum between ionization and density in \textsc{21cmfast} (points) and phenomenological models (lines)} for models 1 (top) and 2 (bottom). As in Fig. \ref{fig:bubble_density}, we show predictions for several different smoothing scales (see \S\ref{sec:bubble_density}), which sets the variance in the density field and thus density contour containing $Q$\% of the volume. We have assumed the log-normal BSD, with $R(z)$ and $\sigma(z)$ calibrations shown in Fig. \ref{fig:Pxx_v_21cmfast_cal} and \ref{fig:Pxx_v_21cmfast_evol}.
    }
\label{fig:Pxd}
\end{center}
\end{figure*}

In Fig. \ref{fig:Pxd}, we move on to the phenomenological model's predictions for the cross spectrum between the ionization and density fields. Line-style conventions are the same as in Fig. \ref{fig:bubble_density}. As in the case of the ionization power spectrum, the phenomenological model provides a very reasonable prediction for the \textit{shape} of $k^3 P_{x\delta}$ as a function of $k$. However, there are systematic offsets from the \textsc{21cmfast} models that vary as a function of $\Rsm$. Interestingly, the best matches occur when $\Rsm$ is tied to the peak in $V dn/dR$ (dashed) or the decline of the one-bubble term (dotted).

Though setting $\Rsm$ to the scale where $V dn/d\log R$ peaks is not the obvious choice based on Figures \ref{fig:bubble_density} and \ref{fig:Pxd}, it is the only option that keeps the 21-cm power spectrum positive (see \S\ref{sec:crossterms_2pt}). Because our two-zone IGM model over-estimates the strength of the fluctuations in the ionization field, as well as cross-terms involving ionization and density, \textit{under}-estimating the density of ionized gas acts as a countermeasure that keeps the amplitude of fluctuations in check. While this is far from an ideal solution, for our purposes it may not matter, as long as our predictions for the 21-cm fluctuations are reasonably accurate.

Having compared predictions for each component of our model to the statistics of the ionization field and its relation to the density field, we now do one last comparison to the 21-cm power spectrum. With $R$ and $\sigma$ calibrated to \textsc{21cmfast} models at a series of $Q$ values, we can perform a one-dimensional fit to the 21-cm power spectrum produced by \textsc{21cmfast} varying $\Ts$ alone, and see if the $\Ts$ found in our fits agrees well with the mean $\Ts$ drawn from \textsc{21cmfast}. Because $\Ts$ in the phenomenological model refers to the mean temperature of the bulk IGM only i.e., neglecting ionized regions), we average over all voxels with $\xHI \geq 0.95$ in the \textsc{21cmfast} boxes \citep[as in, e.g.,][]{HERA2021Theory}. 

\begin{figure*}
\begin{center}
\includegraphics[width=0.98\textwidth]{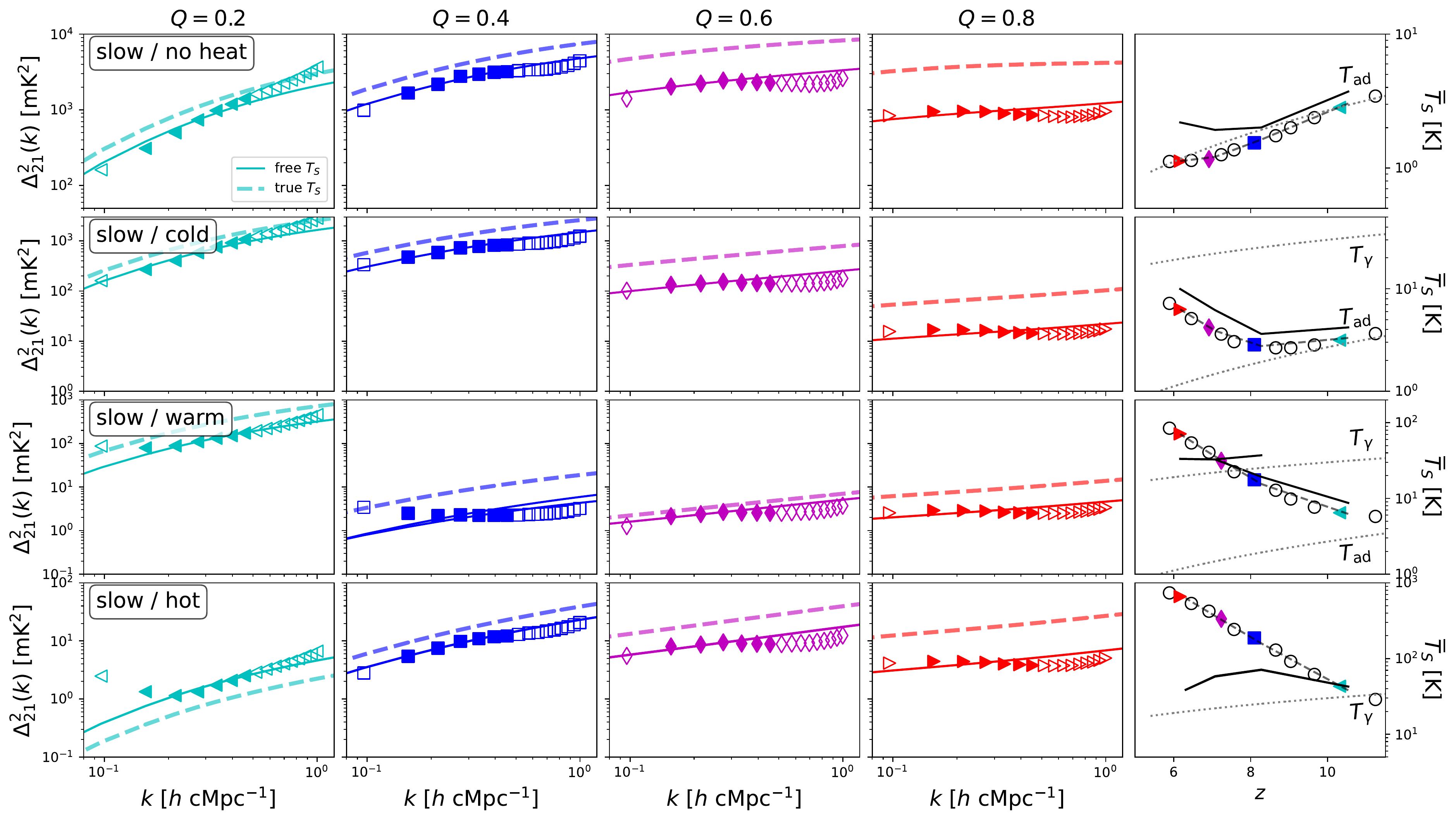}
    \caption{{\bf 21-cm power spectrum from \textsc{21cmfast} (points) and phenomenological models (lines) for the ``slow'' reionization scenario, including several models with increasingly efficient heating (top to bottom).} The first four columns show 21-cm power spectra at $Q=0.2, 0.4, 0.6$, and 0.8, while the final column shows the redshift evolution of $\TS$. In each panel, we freeze $Q=1-\xHIavg$, and take $R$ and $\sigma$ to their calibration values (derived from the ionization power spectrum alone; see Fig. \ref{fig:Pxx_v_21cmfast_cal} and \ref{fig:Pxx_v_21cmfast_evol}). We then fit the 21-cm power spectrum with a single free parameter, $\TS$. The results of this fit are shown as solid lines, while dashed lines adopt the true $\TS$ from \textsc{21cmfast}. Note that because $\TS$ is obtained at each snapshot independently, it can be double-valued (final column), since $\Delta_{21}^2 \propto (1 - \TR/\TS)^2$. Such behaviour does not occur when jointly fitting measurements at different redshifts with a parametric form for $\Ts(z)$ (see \S\ref{sec:21cmfast_fits}).
    }
\label{fig:P21_v_21cmfast}
\end{center}
\end{figure*}

\begin{figure*}
\begin{center}
\includegraphics[width=0.98\textwidth]{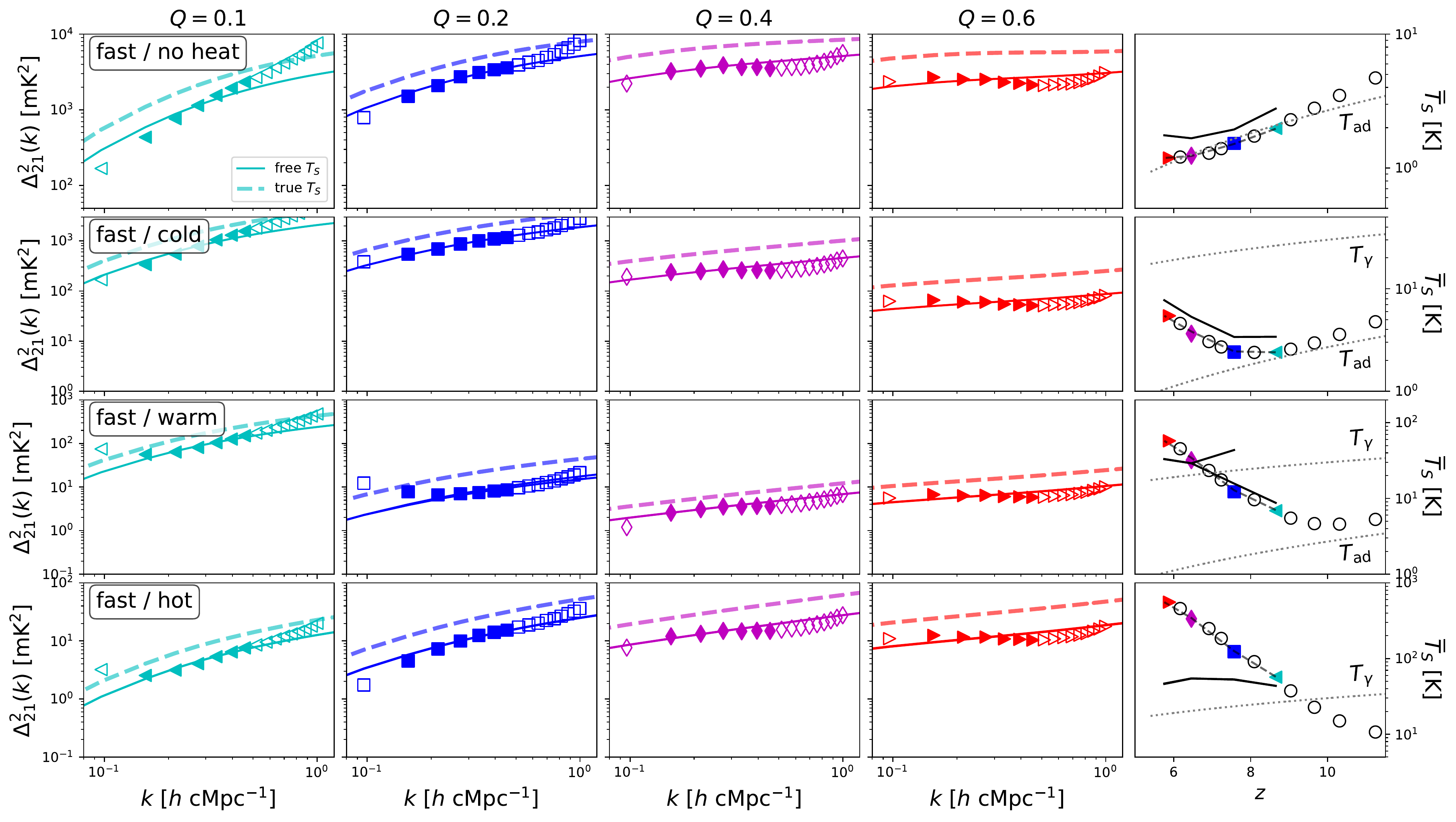}
    \caption{Same as Fig. \ref{fig:P21_v_21cmfast}, but for the ``fast'' reionization scenario. The first four columns show 21-cm power spectra at $Q=0.1, 0.2, 0.4$, and 0.6, while the final column shows the redshift evolution of $\TS$. The efficiency of X-ray production increases from top to bottom.}
\label{fig:P21_v_21cmfast_Mturn_1}
\end{center}
\end{figure*}

The results of this test are shown in Fig. \ref{fig:P21_v_21cmfast} and \ref{fig:P21_v_21cmfast_Mturn_1}, for slow and fast reionization scenarios, respectively. In each case, the model in the first row has minimal heating from astrophysical sources, while subsequent rows have increasingly efficient heating. Dashed lines in the first four columns show 21-cm power spectra obtained using the phenomenological model \textit{when assuming the true underlying value for $T_S$}. Solid lines instead show power spectra obtained if $T_S$ is allowed to vary freely in a 1-D fit. The final column shows the temperature evolution, again including the true evolution (points and dashed line) and that recovered in the 1-D fit to the power spectrum (solid lines).

There is much to unpack from Figs.~\ref{fig:P21_v_21cmfast} and \ref{fig:P21_v_21cmfast_Mturn_1}. Generally speaking, the phenomenological model performs best in the early stages of reionization, $Q \lesssim 0.6$ (first three columns), and begins to struggle later in reionization, particularly when the spin temperature is large. This result is not unexpected; the sharp spherical bubbles in our model will result in stronger ionization fluctuations at fixed $Q$ compared to \textsc{21cmfast}, meaning we should expect to underestimate the contrast, $1-\Tcmb/\Ts$. Indeed, this is generally what we find in Figs.~\ref{fig:P21_v_21cmfast} and \ref{fig:P21_v_21cmfast_Mturn_1}. The shape of our power spectra are a reasonably good match, becoming flatter near the midpoint of reionization. Similarly, the shape of our thermal histories are in good agreement with the \textsc{21cmfast} models, though exhibit small biases, as described above.

There are cases in which the 21-cm power on large scales departs from the phenomenological model's prediction even at early times. We have verified that this excess power on large scales $k \lesssim 0.1 \ \invMpch$ and early times $Q \lesssim 0.2$ (see bottom two panels in the first column) is not due to the presence of temperature fluctuations by running an additional \textsc{21cmfast} simulation in which we set $\Ts = \infty$ everywhere by hand. This large-scale power is likely due to the additional ionization from X-ray sources, which our phenomenological approach is not equipped to model. As we will see in \S\ref{sec:21cmfast_fits}, uncertainties on these $k$ modes are large, and so unlikely to contribute significantly to biasing our fits.

Finally, notice that in some cases, $\Ts$ values just above and just below $\Tcmb$ are both valid, because the power spectrum is proportional to $(1 - \Tcmb/\Ts)^2$ (neglecting adiabatic corrections). When this is the case, we show two solid lines, one for each $\Ts$ solution. In reality, this degeneracy will be broken by fitting data at multiple redshifts simultaneously, unless the true temperature evolution of the IGM is very gradual, and hovers near $\Ts \approx \Tcmb$. However, the phenomenological model is never driven to $\Ts \gg \Tcmb$ -- as seen in the bottom rows of Figs.~\ref{fig:P21_v_21cmfast} and \ref{fig:P21_v_21cmfast_Mturn_1}. Though the spin temperature saturates early in \textsc{21cmfast} in each case, the recovered $\Ts$ by the phenomenological model never exceeds $\sim$ few times $\Tcmb$. This behaviour is not unexpected. First, we are less and less sensitive to the spin temperature as $\Ts$ grows much larger than $\Tcmb$, and so should expect constraints to be poor. And second, given that our model over-estimates the strength of ionization fluctuations at fixed $Q$ due to the idealized assumption of sharp, spherical bubbles, we will then \textit{underestimate} the contrast, $1-\Tcmb/\Ts$ when fitting to a given 21-cm power. As we will see in the next section, uncertainties on the inferred spin temperature history inflat dramatically in these regime, providing an indicator that a measurement lies in a region of parameter space captured poorly by our model in its current form.

In this section, we showed that the phenomenological model provides a reasonably accurate match to predictions from \textsc{21cmfast} simulations. There are differences, as expected, but they are largely systematic, with biases at the level of $\sim 20-40$\% in the amplitude of the ionization power spectrum and ionization--density cross spectrum. These biases will of course affect our ability to recover the mean spin temperature and ionized fraction in fits to mock datasets generated with \textsc{21cmfast}, since these parameters act largely as normalization factors in the phenomenological model, whereas $R$ and $\sigma$ (or $\gamma$) carry more shape information. We will assess this possibility in the next section.

\section{Recovery of IGM Properties from 21-cm Mock Power Spectra} \label{sec:fits}
We now determine the extent to which the mean properties of the IGM and parameters describing the size distribution of bubbles can be recovered from observations using phenomenological models. We explore two different scenarios. First, we fit a mock signal generated with the phenomenological model itself (\S\ref{sec:selffit}), and explore the power spectrum's sensitivity to the BSD in more detail in \S\ref{sec:bsd_pca}. Then, in \S\ref{sec:21cmfast_fits}, we fit mocks generated with \textsc{21cmfast}. We use \textsc{emcee} \citep{ForemanMackey2013} (version 2.2.1), an implementation of the affine-invariant sampler of \cite{Goodman2010}, to perform all Markov Chain Monte-Carlo (MCMC) fits to mock observations.

\subsection{Fits to phenomenological model mocks} \label{sec:selffit}
We begin with a simple exercise to make sure that the parameters of the the phenomenological model can be recovered under idealized circumstances. The input model adopted for this calculation assumes $Q=0.4$, $\TS=3$ K, $R_b=5$ cMpc, and $\sigma_b=1$ (i.e., a log-normal BSD). We compute error-bars appropriate for HERA using \textsc{21cmsense}\footnote{\url{https://github.com/jpober/21cmSense}}~\citep{Pober2013,Pober2014} under the assumption of ``moderate'' foregrounds, and set a superhorizon buffer of $a=0.05 \ \invMpch$.
We take a system temperature
\begin{equation}
    T_{\rm sys} = 100\,\mathrm K + 120\,\mathrm K  \times(\nu/150\rm GHz)^{-2.55},
\end{equation}
following~\cite{DeBoer2017}, and assume 1 year (1080 hours) of observation time.
We bin linearly in wavenumbers (with $\Delta k=0.1 \ \invMpch$), and frequency (with a bandwidth $\Delta \nu=8$ MHz), from $z \approx 6-24$.
This results in a signal-to-noise ratio SNR $\sim 200$ for the EOS21 model (1d in Table~\ref{tab:21cmfast}).

The noise on the 21-cm power spectrum has a cosmic-variance component that depends on the fiducial signal that we study.
Rather than re-running {\tt 21cmSense} for each different model in our array, we have separated the thermal $\sigma_{\rm th}$ and cosmic-variance $\sigma_{\rm CV}$ contributions to the error, which can be added to find the full error as \citep{Munoz2021ETHOS}
\begin{equation}
    \sigma_{\rm full} = \sigma_{\rm th} + \sigma_{\rm CV},
\end{equation}
where $\sigma_{\rm th}$ is a standard output of {\tt 21cmSense}, and
\begin{equation}
     \sigma_{\rm CV} = a_{21} \Delta^2_{21}
\end{equation}
is given by the $k$- and $z$-dependent coefficients $a_{21}$, which we find for our chosen bins.
This relationship simply states that the total noise $\sigma_{\rm full}$ grows linearly with the size of the fiducial signal, due to cosmic variance\footnote{As a point of comparison, the variance of the CMB temperature power spectrum $C_\ell$ scales as $\sigma^2(C_\ell) \propto (C_\ell + N_\ell)^2$~\citet{Kamionkowski:1996ks}, scaling the same way as our result here.}.

\begin{figure*}
\begin{center}
\includegraphics[width=0.49\textwidth]{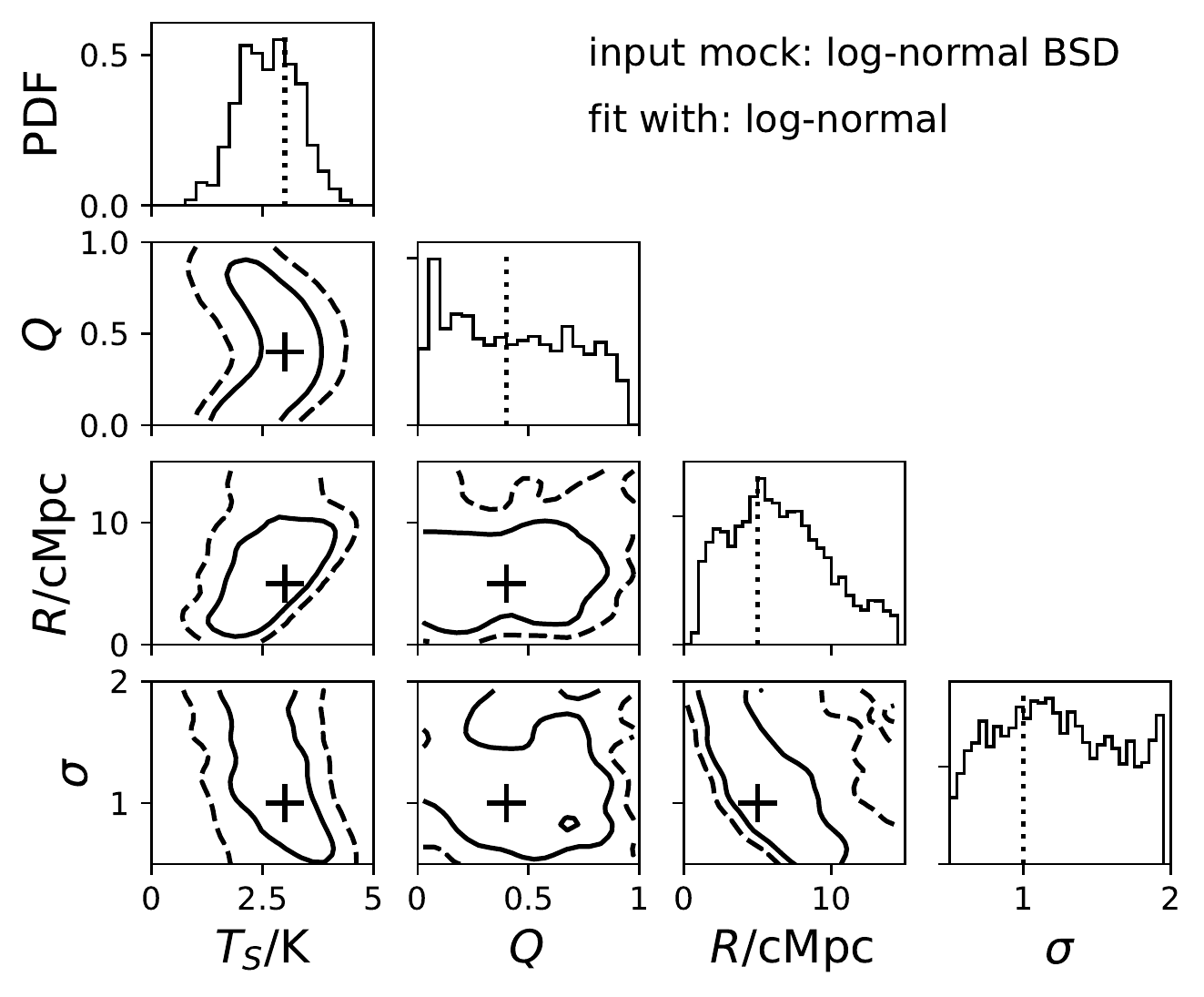}
\includegraphics[width=0.49\textwidth]{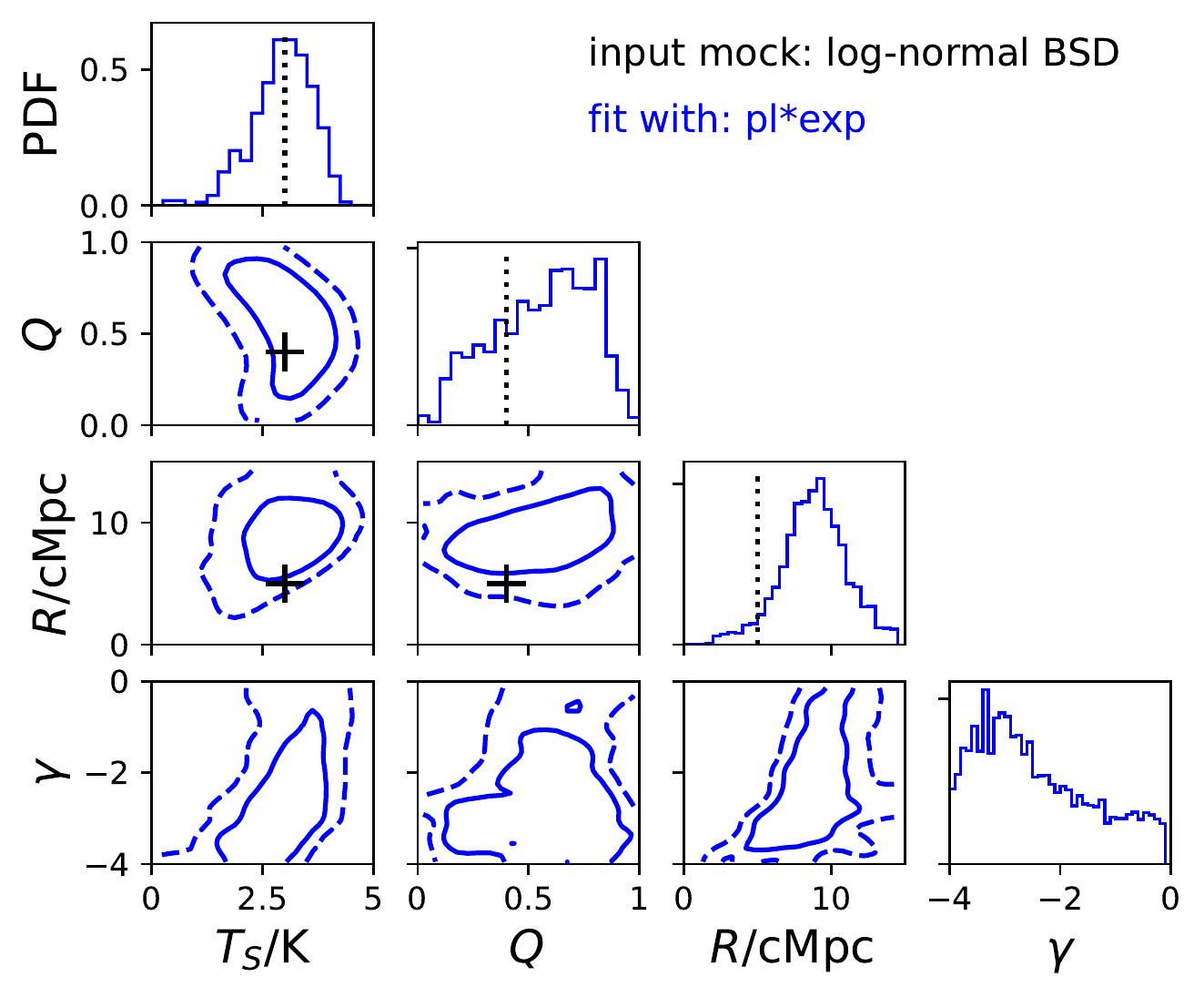}
    \caption{{\bf Parameter constraints obtained via fits to a mock phenomenological signal with $\Ts=3$ K, $Q=0.4$, $R=5$, and $\sigma=1$ at $z=8$ using both BSD parameterizations.} \textit{Left:} Posterior distributions when mock signal is fit with the same BSD as that used to generate the mock (lognormal). \textit{Right:} Results when we assume a power-law-times-exponential BSD in the fit. Solid and dashed curves indicate 68 and 95\% confidence regions, respectively, while the input values are denoted by crosses in the interior panels and vertical dotted lines for the 1-D posteriors. Only the spin temperature is constrained well in each case (first column), though the 2-D relationship between $\TS$ and $R$ can also be reliably constrained.}
\label{fig:mock_rec_self}
\end{center}
\end{figure*}

In Figure \ref{fig:mock_rec_self}, we show the result of this simple forecast. In the left triangle plot, we show the recovery of the log-normal BSD model parameters, and in the right panel, we show the results of a fit that uses a power-law-times-exponential BSD, though the input mock remains that generated with a log-normal BSD. In the first case, the true input values are recovered well, though generally with large uncertainties. For example, $Q$ can only be constrained to $0.04 \lesssim Q \lesssim 0.91$ at $2 \sigma$ confidence. The typical bubble size is constrained to $3.5 \lesssim R / \rm{cMpc} \lesssim 12.8$ (1$\sigma$), with no real constraints on $\sigma$. The mean spin temperature of the IGM, however, is constrained well, to $\Ts \simeq 3 \pm 1$ K. The constraint on the spin temperature largely holds even if fitting with the ``wrong'' BSD, as shown in the right panel of Fig. \ref{fig:mock_rec_self}. The ionized fraction $Q$ is similarly only weakly constrained, while the typical bubble size is biased to slightly larger sizes, $R \simeq 8 \ \rm{cMpc}$. In principle, one could perform model selection and select the best-fitting parameterization with the Bayesian evidence, though we defer such an analysis to future work.

The weakness of a single-epoch fit is not unexpected. Our model has four parameters, and is being used to fit a curve that only mildly departs from a pure power-law. We chose an input mock with a cold enough spin temperature that only solutions with $\Ts \ll \Tcmb$ are viable, which significantly helps to constrain $\Ts$. The results are largely consistent for each BSD, in that $\Ts$ is the best constrained parameter, with the others only weakly constrained. We take a deeper look at the power spectrum's sensitivity to the BSD next.

\subsection{Sensitivity of PS to BSD} \label{sec:bsd_pca}
In \S\ref{sec:selffit} we emphasized potential constraints on $Q$ and $\Ts$, and discussed the BSD only as a nuisance and its potentially odd behavior in the absence of physical priors. However, given that our phenomenological model takes BSDs as input for power-spectrum predictions, a complementary approach would be to examine the extent to which power spectrum measurements can place constraints on BSDs.

\begin{figure}
\begin{center}
\includegraphics[width=0.45\textwidth]{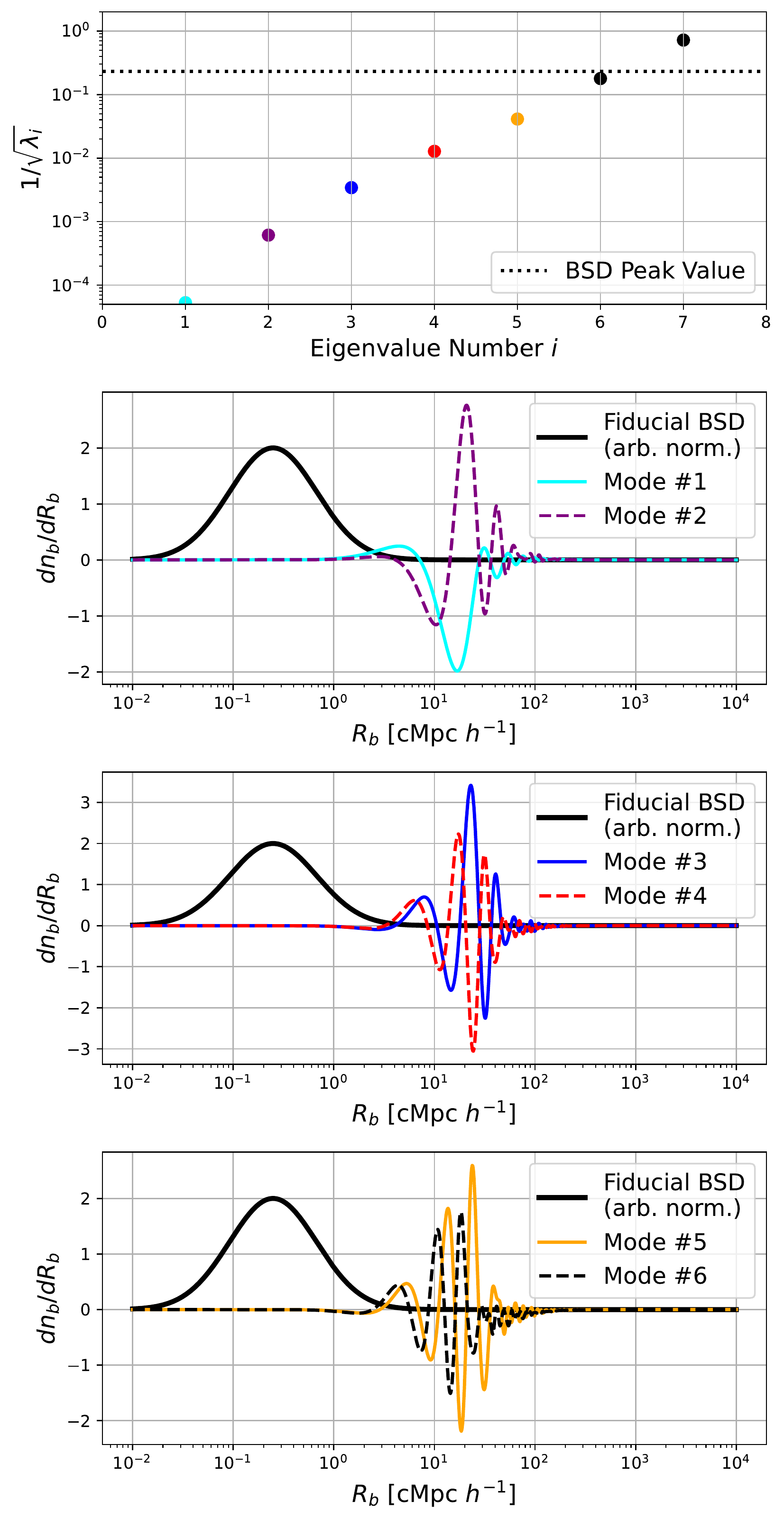}
    \caption{{\bf Principal components modes of accessible modes to power-spectrum constraints on the bubble size distribution (BSD).} \emph{Top:} Inverse square root of the eigenvalue spectrum of the principal components, which provides a sense for the measurability of each principal component. \emph{Next three:} First six principal components of the BSD, representing the most measurable perturbations on a fiducial log-normal bubble BSD (solid black). Note that in contrast to Fig.~\ref{fig:bsd}, here we plot $dn_b /dR_b$ rather than $Q^{-1} V_b dn_b / d\log R_b$.}
\label{fig:modes}
\end{center}
\end{figure}

Consider a BSD that is given by our fiducial log-normal form plus an arbitrary perturbation in each $\ln(R_b)$ bin. Such a model possesses a large number of free parameters -- one perturbation amplitude per discrete $\ln (R_b)$ bin -- and thus the expectation is that a single power spectrum will not likely be able to place meaningful constraints on every degree of freedom of the BSD. Given this limitation, we instead perform a principal component analysis to quantify a few shapes in the BSD that can be easily probed using power spectrum observations. As our starting point, we compute a Fisher matrix to quantify the information content on each perturbation amplitude that is contained in the power spectrum. This is given by
\begin{equation}
F_{\alpha \beta} = \sum_{i} \frac{1}{\sigma_{{\rm full},i}^2} \frac{\partial \Delta^2(k_i)}{\partial \eta_\alpha} \frac{\partial \Delta^2(k_i)}{\partial \eta_\beta},
\end{equation}
where $\sigma_{{\rm full},i}$ is the error bar on a measurement of the $i$th bin of the power spectrum (as described in \S\ref{sec:selffit}) and $\eta_\alpha$ denotes the value of $dn_b / dR_b$ in the $\alpha$th $\ln (R_b)$ bin. It is understood that the derivatives are to be evaluated at a set of fiducial values for any free parameters in the model. Performing an eigen decomposition of this Fisher matrix provides a set of eigenvectors that serves as a series of orthonormal basis templates for the BSD. Figure \ref{fig:modes} provides some example eigenvectors and eigenvalues. The top panel shows the inverse square root of the eigenvalues, which quantify the error in a potential inference of the amplitude of each template from a power spectrum measurement. The templates are therefore ordered in the sense that they represent a set of modes ordered from most precisely measurable to least precisely measurable. The corresponding eigenvectors are shown in the bottom three panels, along with the fiducial BSD. Note that in this section, we plot $dn_b /dR_b$ rather than $Q^{-1} V_b dn_b / d\log R_b$, since it is the former that is an input in the \textsc{micro21cm} package. Since the eigenvectors are normalized, the errors in the top panel can be directly compared to the typical amplitude of $dn_b /dR_b$. The dotted horizontal line in the top panel indicates the peak value of the BSD, providing a rough sense for what modes might be measurable.

Several trends are immediately apparent. First, we note that all the eigenvectors are relatively localized to the high end of the BSD. Unsurprisingly, the scales over which these template modes have appreciable amplitude is in rough correspondence to the scales that we probe using the power spectrum. Second, we see that the modes are essentially tapered Fourier-like modes. Each mode measures finer and finer details of the upper end of the BSD. However, from the top panel we see that only the first few are likely to be measurable---far fewer than the $1000$ bins in $\ln (R_b)$ that were perturbable in our analysis. We thus conclude that while the power spectrum certainly does have some sensitivity to the BSD, the information that can be extracted about it may be limited. Of course, these results do depend on the fiducial BSD used in the analysis, and we find from our experimentation that although our qualitative trends seem fairly robust, the details can vary substantially.

\subsection{Fits to \textsc{21cmfast} mocks} \label{sec:21cmfast_fits}
We now test the phenomenological model's ability to recover the mean IGM properties of mocks generated with \textsc{21cmfast}. While semi-numerical models are known to produce 21-cm power spectra in reasonably good agreement with radiative transfer simulations, at least in the saturated limit \citep{Zahn2011,Hutter2018}, to our knowledge, there has yet to be an attempt to compare analytic and semi-numerical models of reionization or to recover the inputs of one model from the outputs of another. Once again, we expect the phenomenological model to struggle most when temperature fluctuations or partial ionization are important in the input \textsc{21cmfast} mocks, and/or late in reionization when the BSD is an increasingly poor descriptor of the ionization field. We plan to experiment with a non-uniform temperature field in the future, as there is evidence that semi-analytic and semi-numerical techniques agree fairly well before reionization \citep[see][]{Schneider2021}. However, for now we proceed with our model as-is, in order to establish a baseline for potential development in the future.

As in \S\ref{sec:selffit} and \S\ref{sec:bsd_pca}, we use uncertainties generated with \textsc{21cmsense} for HERA with moderate foregrounds, which provides a fiducial benchmark for the state of 21-cm observations over the next few years, and explore all eight \textsc{21cmfast} models introduced thus far (in \S\ref{sec:21cmfast_mocks} and Fig. \ref{fig:mocks_21cmfast}): four different X-ray heating scenarios for each reionization scenario. We fit all redshift snapshots between $6 \lesssim z \lesssim 10$, including only modes in the range $0.1 \leq k / [\invMpch] \leq 1$.

One could in principle let $R$, $\sigma$, $Q$, and $\Ts$ vary independently at every redshift of a multi-epoch fit, though this will of course result in a very high-dimensional model. In addition, single-epoch fits only loosely constrain the ionized fraction (see Fig. \ref{fig:mock_rec_self}), since they lack even simple priors, e.g., that $Q$ increase from high redshift to low. Motivated by Fig. \ref{fig:Pxx_v_21cmfast_evol}, we employ simple functions for the evolution of each parameter in subsequent MCMC fits. For $R$, we employ a power-law in $Q$, which we found performed more efficiently than a a power-law in $z$. We further parameterize $Q$ and $\Ts$ as power laws in redshift and assume $\sigma$ is a constant, which results in a total of 7 free parameters. This allows us to compare to \textsc{21cmfast} models over a range of redshifts, and fit to multi-epoch mock datasets without dramatically increasing the dimensionality of the fit.

In the future, large databases of semi-numeric models (e.g., Prelogovic et al, in prep.), could be used to map out a prior volume in ($R$, $\sigma$, $Q$) space in order to avoid unphysical regions and so reduce uncertainties on $Q$ and $\Ts$. Here, however, we assume broad, uninformative priors on each parameter. We take:
\begin{itemize}
    \item $Q \in [0,1]$ and $d\log Q / d\log z = [-20,0]$.
    \item $\log_{10} \TS /\rm{K} \in [-1, 3]$ and $d\log \TS / d\log z \in [-20,5]$.
    \item $R(Q=0.5) /\rm{cMpc} \in [0.5, 50]$ and $d\log R / d\log Q \in [0, 5]$.
    \item $\sigma \in [0.25, 2.5]$.
\end{itemize}
We also enforce $Q \geq 0.99$ at $z=5.3$, a conservative end-of-reionization prior consistent with the latest interpretation of Ly-$\alpha$ forest constraints \citep{Becker2015,Bosman2021,Keating2020}.

Before we show the results of these fits, we note that each reionization scenario spans a different range in \textit{ionized fraction} over the same interval in \textit{redshift} used for fitting. For example, the IGM in the ``slow'' reionization model is already $\simeq 30$\% ionized by $z \sim 10$, the highest redshift used in the fit, while the ``fast'' reionization scenario has $Q \lesssim 0.1$ at $z \simeq 10$ (see Fig. \ref{fig:mocks_21cmfast}). As a result, we may see different outcomes in the fits given that our model fares poorly at the very end of reionization, and there are effectively more data points late in reionization for the ``slow'' scenario.


\begin{figure*}
\begin{center}
\includegraphics[width=0.98\textwidth]{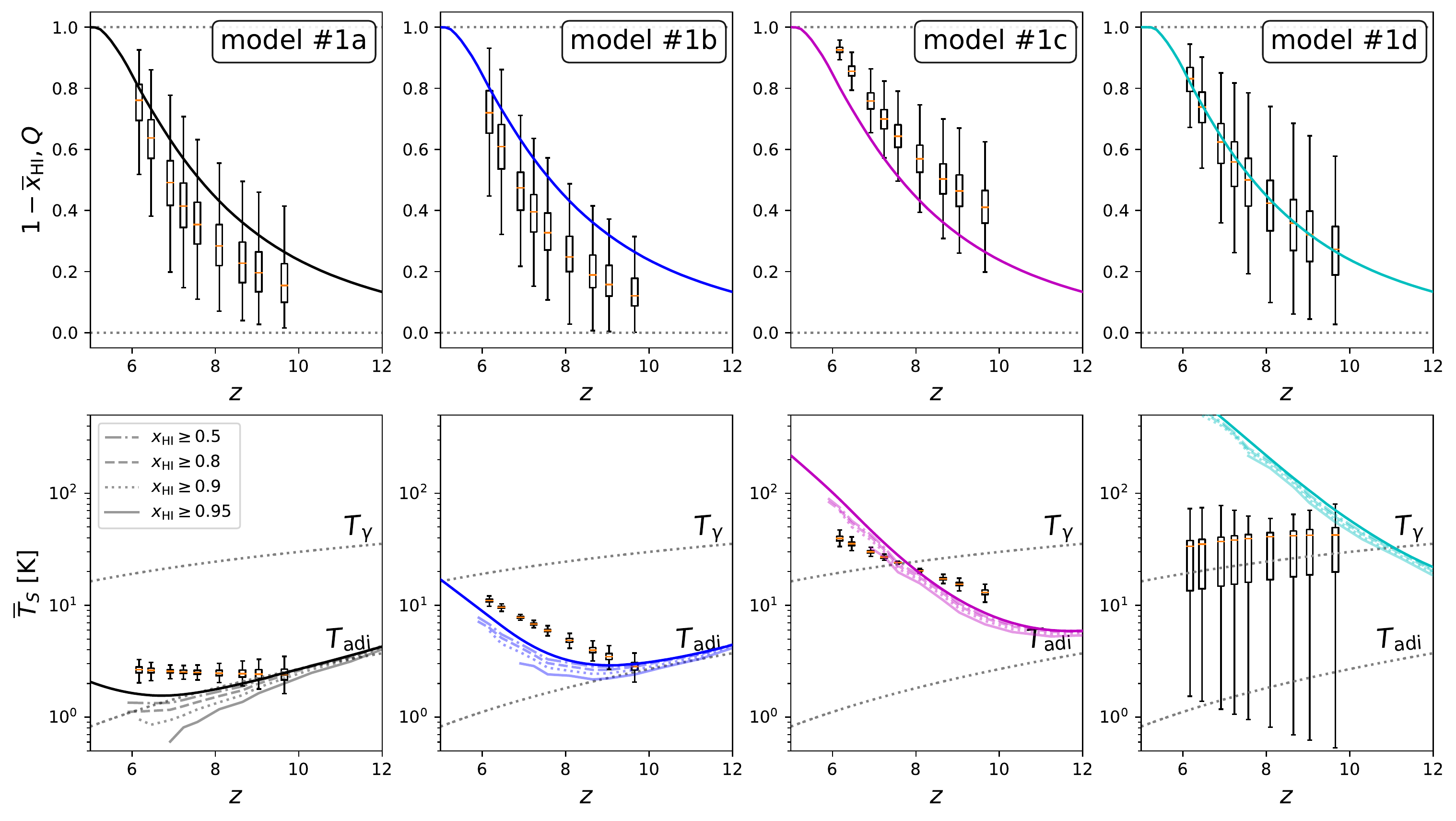}
    \caption{{\bf Recovery of mean ionization (top) and spin temperature (bottom) histories from \textsc{21cmfast} mocks,} for each ``slow'' reionization model. Lines show true evolution extracted from \textsc{21cmfast} directly, while recoveries are shown with error-bars, the full extent of which indicate the 95\% confidence region, and the boxes the 68\% confidence interval. Linestyles in the bottom row indicate the spin temperature averaged over voxels above different neutral fraction thresholds, as indicated in the legend. There are clear biases in both quantities, though the thermal histories are constrained well enough to associate each mock with the correct level of X-ray heating (see \S\ref{sec:21cmfast_fits}).
    }
\label{fig:21cmfast_recon}
\end{center}
\end{figure*}

\begin{figure*}
\begin{center}
\includegraphics[width=0.98\textwidth]{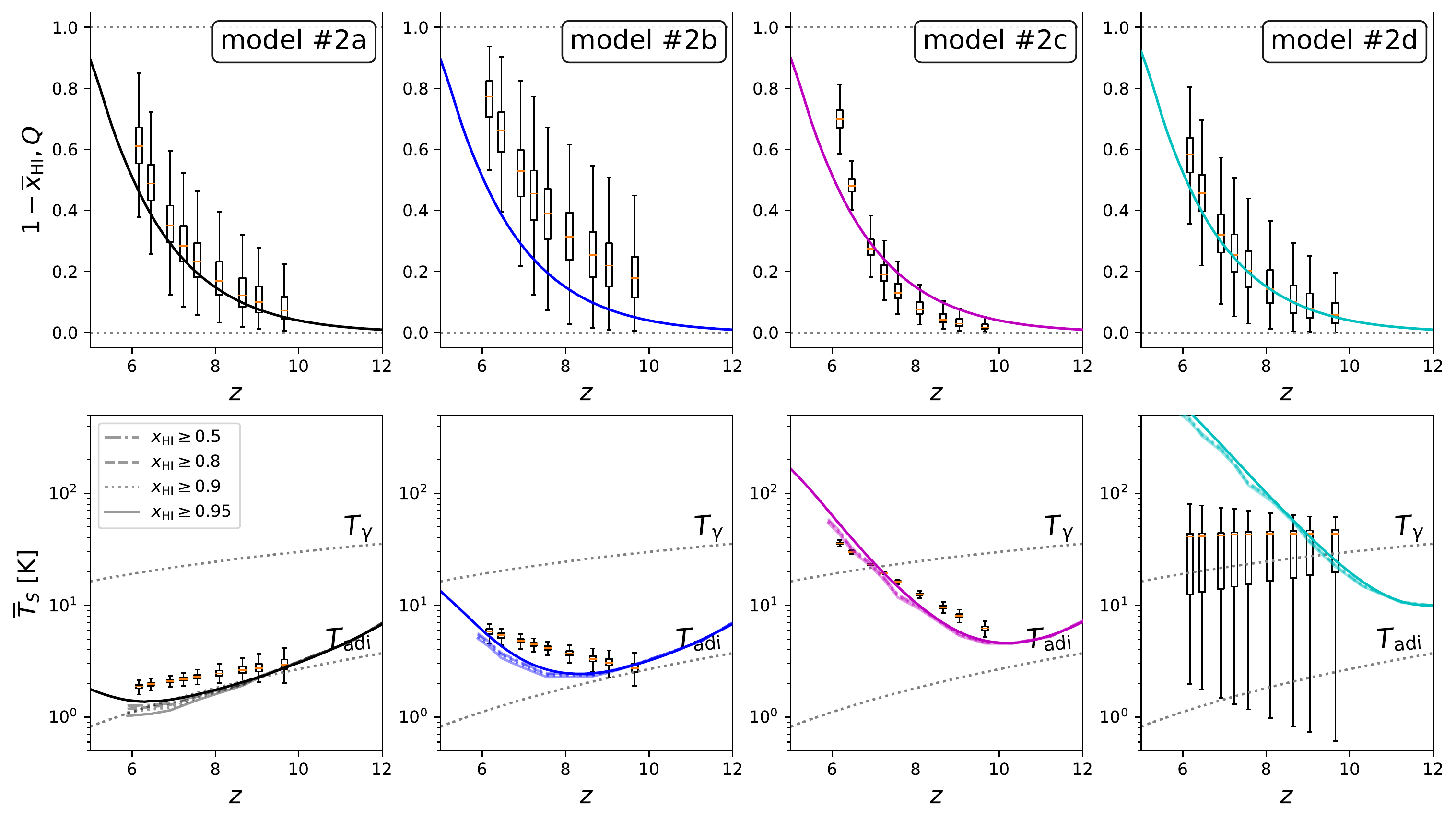}
    \caption{{\bf Recovery of mean ionization (top) and spin temperature (bottom) histories from \textsc{21cmfast} mocks,} for each ``fast'' reionization model. Lines show true evolution extracted from \textsc{21cmfast} directly, while recoveries are shown with error-bars, the full extent of which indicate the 95\% confidence region, and the boxes the 68\% confidence interval. Linestyles in the bottom row indicate the spin temperature averaged over voxels above different neutral fraction thresholds, as indicated in the legend. Though some biases are visible in both quantities, the early history of reionization is better constrained than that of the ``slow'' scenario (Fig. \ref{fig:21cmfast_recon}), likely because the first half of reionization lies in the fit range, $6 \lesssim z \lesssim 10$. Once again, the thermal histories are biased, but are constrained well enough to associate each mock with the correct level of X-ray heating (see \S\ref{sec:21cmfast_fits}).}
\label{fig:21cmfast_recon_Mturn}
\end{center}
\end{figure*}

In Fig. \ref{fig:21cmfast_recon} and \ref{fig:21cmfast_recon_Mturn}, we show our recovery of the mean ionization history (top) and spin temperature history (bottom), for ``slow'' and ``fast'' reionization models, respectively. In each plot, we show results for increasingly efficient X-ray heating going from left to right.

Starting first with the ``slow'' models, two things are clear immediately at a glance: (i) the recovered ionization histories have large uncertainties and are generally biased low, and (ii) the recovered thermal histories, while also slightly biased, are correct in order of $L_X/\rm{SFR}$, and so are good enough to identify the appropriate heating scenario. However, for the most efficient X-ray heating scenario (model 1d; right-most columns), uncertainties on $\Ts$ grow much larger, as we are increasingly insensitive to $\Ts$ once it becomes large. This is in some sense a good thing, i.e., huge errors on $\Ts$ but \textit{not} on $Q$ may alone indicate $\Ts \gg \Tcmb$.

The ``fast'' reionization scenario in Fig. \ref{fig:21cmfast_recon_Mturn} shows the same general trends in its recovery of the thermal history. However, here, the early stages of reionization are recovered much more accurately, while it is the later stages that suffer from more of a bias. This is at least in part caused by our decision to fit both models over the range $6 \lesssim z \lesssim 10$. For the ``fast'' models, the full rise in 21-cm power from $Q \simeq 0$ to $Q \simeq 0.5$ lies within the fitted range, but the latter half of reionization does not. Because the mock contains the part of reionization that the model predicts best, a more reliable fit is obtained. In contrast, the ``slow'' scenario covers the middle $\sim 50$\% of reionization, so the fit cannot leverage the model's accuracy at $Q \lesssim 0.3$.

There are other factors at play, aside from the fitted redshift range, that could explain the biases in Fig. \ref{fig:21cmfast_recon} and \ref{fig:21cmfast_recon_Mturn}. For example, as discussed previously, because our model over-estimates ionization fluctuations we should expect to underestimate the contrast. This is clearly the case in the scenarios with minimal X-ray heating -- the recovered $\TS$ is higher than the true $\TS$ once reionization gets underway. In the final scenario, with very efficient X-ray heating, the exact temperature is constrained very poorly (last column). However, once the temperature is $\Ts \gg \Tcmb$, the 21-cm background is of course increasingly insensitive to $\Ts$. The bias in our recovery is in part caused once again by our over-estimation of ionization fluctuations, except now, in the emission regime, $\Ts$ must be reduced to preserve large-scale 21-cm power as $Q$ rises.

The other likely source of error is our neglect of temperature fluctuations and partial ionization. While our models a and d are nearly equivalent to cases with identically zero heating and full saturation, respectively, the intermediate cases b and c likely have residual temperature fluctuations to some extent during reionization. Similarly, cases c and d are most likely to have partial ionization caused by strong X-ray backgrounds. These factors could be causing the change in the recovered ionization histories as a function of $L_X/\rm{SFR}$. We do not attempt to quantify the magnitude of these effects here, but a closer look may be warranted in future studies.



\begin{figure*}
\begin{center}
\includegraphics[width=0.98\textwidth]{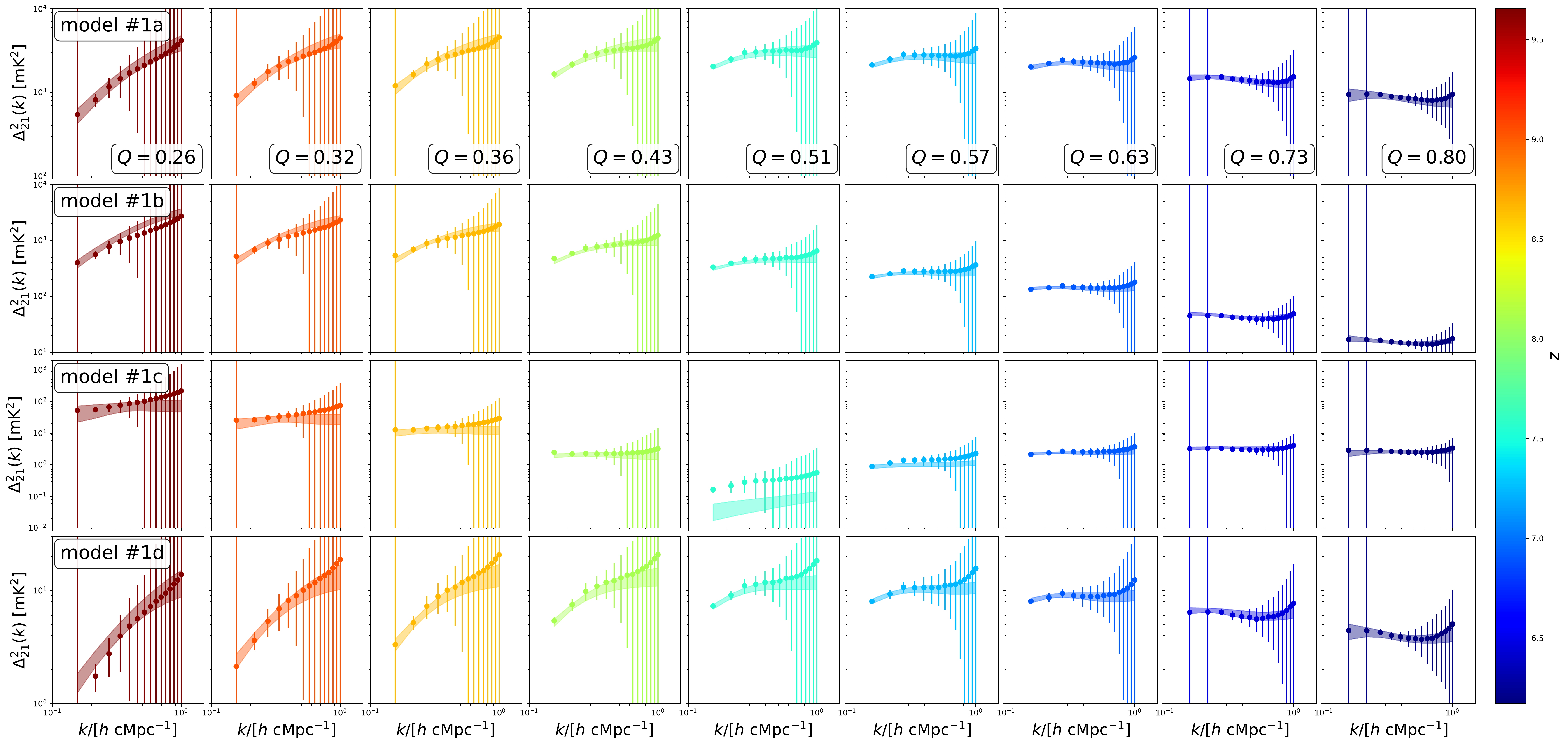}
    \caption{{\bf Recovered 21-cm power spectra from fits to ``slow'' reionization \textsc{21cmfast} mocks.} From top to bottom, we show models with increasing $L_X/\rm{SFR}$, in order of increasing global ionized fraction from left to right. Shaded regions show 68\% confidence reconstructions, while solid lines indicate the maximum likelihood model.}
\label{fig:21cmfast_recon_ps}
\end{center}
\end{figure*}

For completeness, we show example recovered 21-cm power spectra in Fig. \ref{fig:21cmfast_recon_ps}. Now, the four different ``slow'' reionization scenarios are shown from top to bottom, in order of increasing X-ray heating, at nine different mean ionized fractions from left to right. In most cases, the power spectrum is recovered very well, indicating that our 7-parameter model is sufficiently flexible to capture both the time evolution and shape of the 21-cm power spectrum.

\section{Discussion \& Conclusions} \label{sec:conclusions}
In this work, we have introduced a simple, phenomenological model for the 21-cm power spectrum during reionization that abstracts away assumptions about galaxy formation and instead works directly in terms of the bubble size distribution, mean ionized fraction, $Q$, and mean spin temperature, $\TS$. The goal was to build intuition for the results of more sophisticated models like \textsc{21cmfast}, and determine if efficient, IGM-focused models like ours may be sufficiently accurate for inference in the next few years, as upper limits from, e.g., MWA, HERA, and LOFAR continue to improve. To our knowledge, this is the first attempt to recover the IGM properties from 21-cm power spectra generated with \textsc{21cmfast} using a completely different model \citep[though see][for a similarly-motivated analysis in the 21-cm imaging context]{Zhou2021}.

We find that a binary ionization field, with a size distribution of bubbles given by simple log-normal and power-law-times-exponential functions, gives rise to 21-cm power spectra in qualitative agreement with those generated by semi-numeric models (see Fig. \ref{fig:bsd} and \ref{fig:components}). We provide a grid of model predictions for the large-scale 21-cm power spectrum at $z=8$ in Fig. \ref{fig:grid2d}, which show that ``cold reionization'' models, with $\Ts \simeq 1.8$ K (the limit corresponding to pure adiabatic cooling), can be disfavoured by upper limits at the $\sim 10^3 \ \rm{mK}^2$ level.

Indeed, the recent \citet{HERA2021} upper limits ($(30.76 \ \rm{mK})^2$ at $z \sim 7.9$ and $k=0.192 \ \invMpch$) imply spin temperatures in excess of the adiabatic limit, as found in four independent analyses, including an analytic bias model, our phenomenological model, and two semi-numerical models \citep{HERA2021Theory}. Our model predicts that an order-of-magnitude improvement, resulting in upper limits of $\sim 10^2 \ \rm{mK}^2$ at $z\simeq 8$, would drive lower limits on the spin temperature securely to $\Ts \gtrsim 10$ K. This is an important milestone, as $\Ts \sim 10$ K is  expected in models anchored to galaxy luminosity functions \citep{Mirocha2017,Park2019}, if one assumes there is no evolution in the efficiency of X-ray production \citep[see also Fig. 7-8 in][]{HERA2021Theory}. As a result, $\sim 100 \ \rm{mK}^2$ limits would substantiate expectations of redshift evolution in the $L_X$--SFR relation due to declining metallicities in galaxies at high redshift \citep{Fragos2013,Brorby2016}.

We also performed several forecasts, first a single-epoch fit to a mock 21-cm power spectrum generated with the phenomenological model (\S\ref{sec:selffit}), followed by a multi-epoch fit to mocks generated with \textsc{21cmfast} (\S\ref{sec:21cmfast_fits}). 

When fitting mock signals generated with the phenomenological model, we do recover the input model parameters, though uncertainties are generally large (see Fig. \ref{fig:mock_rec_self}). Simultaneously fitting data at multiple redshifts is thus vital to obtaining tight constraints on model parameters, e.g., the mean ionized fraction. The IGM spin temperature, $\Ts$, is an exception here, at least for strong signals. The detailed shape of the BSD is likely beyond the reach of current experiments, though, encouragingly, the power spectrum is sensitive to the typical bubble size and the distribution of sizes just above the peak (see \S\ref{sec:bsd_pca}).

We also fit a total of eight \textsc{21cmfast} mocks: a ``slow'' and ``fast'' reionization scenario with four different X-ray heating scenarios for each. Because our model \textit{over-estimates} the ionization fluctuations (see Fig. \ref{fig:Pxx_v_21cmfast_cal}-\ref{fig:Pxx_v_21cmfast_evol}), we generally \textit{underestimate the contrast}, $1-\Tcmb/\Ts$. As a result, for cold reionization scenarios, the phenomenological model yields slightly higher temperatures than are assumed by the input mock, and if heating drives $\Ts \gg \Tcmb$, we instead obtain $\Ts$ values that are biased low. In detail, the recovered ionization and thermal histories are  underwhelming in some parts of parameter space (Fig. \ref{fig:21cmfast_recon} and \ref{fig:21cmfast_recon_Mturn}). However, one can reliably place recovered thermal histories into broad categories (no heat and low/medium/high heat), a triumph for such a simple, phenomenological approach.



Finally, we note that our model adopts several key simplifying assumptions, improvements to which may bring our approach into closer agreement with semi-numeric models. For example, we assume:
\begin{itemize}
    \item Bubbles are fully ionized, perfectly spherical, and have perfectly sharp edges, which results in an over-estimate of fluctuations in the ionization field (see \S\ref{sec:21cmfast_mocks} and Fig. \ref{fig:Pxx_v_21cmfast_cal}-\ref{fig:Pxx_v_21cmfast_evol}).
    \item The degree of bubble overlap is estimated (see \S\ref{sec:overlap}) but we do not attempt to correct for overlap effects.
    \item The density field is assumed to mirror the ionization field, which allows a simple approach to computing cross-terms (see \S\ref{sec:crossterms_2pt}-\ref{sec:crossterms_Npt}) and the mean density of ionized regions (see \S\ref{sec:bubble_density} and Fig. \ref{fig:bubble_density}), but results in biases (see Fig. \ref{fig:Pxd}).
    \item Though we focus entirely on ionized bubbles, one can instead treat bubbles of a fixed temperature with the same formalism by changing the parameter $\alpha$ (see \S\ref{sec:methods}). However, we do not allow both kinds of fluctuations to operate simultaneously in this work, and in some cases our ability to recover IGM properties from \textsc{21cmfast} models may suffer as a result.
\end{itemize}
With a run-time of less than a second per redshift, we can afford improvements to the model, even if they come with a non-trivial penalty in computational efficiency. The assumptions listed above may thus be prime targets for improving the model, in an attempt to reduce biases in IGM constraints derived from our phenomenological approach, and perhaps even tighten constraints over a broader range of parameter space (e.g., the $\Ts \gg \Tcmb$ regime). We defer an exploration of potential improvements to future work, and in the meantime welcome additions and/or improvements to the code at \url{https://github.com/mirochaj/micro21cm}. \vspace{10pt}

The authors acknowledge Ad\'{e}lie Gorce, Paul La Plante, Stefan Heimersheim, Anastasia Fialkov, and Yuxiang Qin for helpful feedback and encouragement throughout this effort. This material is based upon work supported by the National Science Foundation under Grant No. 1636646, the Gordon and Betty Moore Foundation, and institutional support from the HERA collaboration partners. HERA is hosted by the South African Radio Astronomy Observatory, which is a facility of the National Research Foundation, an agency of the Department of Science and Technology. JBM is supported by a Clay fellowship at the Smithsonian Astrophysical Observatory. SRF was supported by the National Science Foundation through award AST-1812458. and was directly supported by the NASA Solar System Exploration Research Virtual Institute cooperative agreement number 80ARC017M0006. A.L. acknowledges support from the New Frontiers in Research Fund Exploration grant program, a Natural Sciences and Engineering Research Council of Canada (NSERC) Discovery Grant and a Discovery Launch Supplement, a Fonds de recherche Nature et echnologies Quebec New Academics grant, the Sloan Research Fellowship, the William Dawson Scholarship at McGill, as well as the Canadian Institute for Advanced Research (CIFAR) Azrieli Global Scholars program. Computations were made on the supercomputer Cedar at Simon Fraser University managed by Compute Canada. The operation of this supercomputer is funded by the Canada Foundation for Innovation (CFI).

\textit{Software:} numpy \citep{numpy}, scipy \citep{scipy}, matplotlib \citep{matplotlib}, \textsc{emcee} \citep{ForemanMackey2013}, \textsc{camb} \citep{Lewis2000}.

\textit{Data Availability:} The data underlying this article is available upon request.

\bibliography{references}
\bibliographystyle{mn2e_short}

\end{document}